\documentclass{article}
\usepackage[utf8]{inputenc}
\usepackage{Paper}
\usepackage{optidef}

\title{The Impact of Shared Autonomous Vehicles in Microtransit Systems: A Case Study in Atlanta}

\author[1,2]{Jason~Lu}
\author[1]{Tejas~Santanam}
\author[1]{Hongzhao~Guan\footnote{Corresponding author: \href{mailto:hguan7@gatech.edu}{hguan7@gatech.edu}}}
\author[1]{Connor~Riley}
\author[1]{Meen-Sung Kim}
\author[1]{Anthony~Trasatti}
\author[2]{Neda~Masoud}
\author[1]{Pascal~Van~Hentenryck}

\affil[1]{H. Milton Stewart School of Industrial and Systems Engineering,\protect\\ Georgia Institute of Technology}
\affil[2]{Department of Civil and Environmental Engineering,\protect\\ University of Michigan}
\date{\today}

\begin{document}

\maketitle
\begin{abstract}
\noindent
Microtransit systems represent an enhancement to solve the first- and last-mile problem, integrating traditional rail and bus networks with on-demand shuttles into a flexible, integrated system. This type of demand-responsive transport provides greater accessibility and higher quality levels of service compared to conventional fixed-route transit services. Advances in technology offer further opportunities to enhance microtransit performance. In particular, shared autonomous vehicles (SAVs) have the potential to transform the mobility landscape by enabling more sustainable operations, enhanced user convenience, and greater system reliability. This paper investigates the integration of SAVs in microtransit systems, advancing the technological capabilities of on-demand shuttles. A shuttle dispatching optimization model is enhanced to \revision{incorporate SAV functionalities, including zero response-time delay for incoming requests and dynamic re-routing}. A model predictive control approach is proposed that dynamically rebalances on-demand shuttles towards areas of higher demand without relying on vast historical data. Scenario-driven experiments are conducted using data from the \revision{Metropolitan Atlanta Rapid Transit Authority (MARTA)} Reach microtransit pilot\revision{, implemented through a collaboration between MARTA and the Georgia Institute of Technology}. The results demonstrate that SAVs can elevate both service quality and user experience compared to traditional on-demand shuttles in microtransit systems.

\end{abstract}

\clearpage

\section{Introduction}
\label{sec:intro}

The demand for mobility services has grown at an unprecedented rate with the proliferation of smartphone apps and the rise of the sharing economy \citep{chan2012ridesharing,standing2019implications}. These developments have fueled the rapid expansion of digital ridehailing systems such as Uber or Lyft, which provide convenient door-to-door transportation for mobility users \citep{wang2019ridesourcing}. However, traditional ridehailing services exhibit a few critical drawbacks. In particular, they contribute significantly towards empty miles, and by consequence exacerbate congestion in the road network and \revision{reduce} overall system efficiency \citep{erhardt2019transportation,henao2019impact,wu2021assessing,beojone2021inefficiency}. By contrast, while traditional fixed-route transit services provide more sustainable mobility, their limited accessibility due to the first- and last-mile problem often hinders user adoption \citep{venter2020measuring,sogbe2024first,lu2025ibp,lu2026path}.
 
Microtransit, a demand-responsive transport service, is an integrated mobility solution made up of on-demand services to enhance system efficiency and address the first- and last-mile problem for traditional bus and rail networks.
Modern microtransit systems operated by transit agencies use human-operated on-demand vehicles, thereby limiting the potential efficiency of first- and last-mile services \revision{due to policy and operational constraints}. In particular, transit agencies enforce safety protocols that prohibit drivers \revision{from interacting with dispatch-related mobile applications while en route between stops to reduce the risk of driver distraction} \citep{Hart_2009,griffin2014prevalence}. \revision{This restriction prevents shuttles from being dynamically re-routed to accommodate newly arriving requests during active service, even when such requests could be served with minimal detour.} From a user perspective, this can substantially prolong waiting times, as drivers cannot respond to new trip requests \revision{while en route}. From an operational standpoint, shuttles would incur unnecessary empty miles as shuttles may need to backtrack \revision{along} previously traveled routes to accommodate requests. Furthermore, when shuttles are idling, drivers could experience delayed responses to incoming requests due to inattention, further extending user waiting times. Shared autonomous vehicles (SAVs) present the opportunity to nullify these limitations, as their technologies enable dynamic \revision{re-routing} of shuttles to promptly accommodate online user requests \citep{alonso-mora2017ondemand,bruglieri2024efficiently}. \revision{In addition, SAVs can process requests immediately upon arrival without human response delays, effectively eliminating delays arising from delayed responsiveness to dispatch instructions. Once sufficiently mature and capable of reliable operation, SAVs also have the potential to reduce labor costs by eliminating the need for onboard drivers \citep{bosch2018cost}. Furthermore, because SAVs do not rely on human drivers, they eliminate safety risks associated with driver-related factors, such as fatigue and distraction \citep{fagnant2015preparing}.}

This paper studies the impact of SAVs in microtransit systems and evaluates their impact on system performance. It proposes the \revision{Integrated} Real-Time Dispatch and Control (\revision{IRDC}) framework, which consists of shuttle dispatching, routing, and rebalancing models. \revision{The framework comprises two sequential components. The first is a dispatching component that assigns and routes shuttles to serve user requests through the Real-Time Dial-a-Ride System (RTDARS) framework. Depending on the fleet under consideration, either the baseline RTDARS model is used to capture the operational characteristics of human-driven shuttles or an enhanced RTDARS-SAV model is used to capture the operational capabilities of SAVs. The second is a control component that employs a model predictive control (MPC) framework to rebalance idle shuttles toward areas of higher anticipated demand without relying on extensive historical data. To evaluate the impacts of SAV integration and other operational changes in the microtransit system, scenario-driven experiments were conducted using data from a six-month microtransit pilot implemented through a collaboration between the Metropolitan Atlanta Rapid Transit Authority (MARTA) and the Georgia Institute of Technology. The pilot, called MARTA Reach, served as an On-Demand Multimodal Transit System (ODMTS) prototype \citep{vanhentenryck2023reach}.} ODMTS is an emerging type of microtransit system, where trains and buses operate on fixed schedules while shuttles operate dynamically to serve passengers' first- and last-mile connections. Passengers provide their desired origin and destination locations through a mobile application or via telephone, after which ODMTS provides a route to serve them.

The remainder of the paper is structured as follows. Section~\ref{sec:literature} reviews related work.
Section~\ref{sec:methodology} presents the methodology developed for this study. Section~\ref{sec:counterfactual-experiments} describes the scenario-driven experiments based in the Atlanta metropolitan area and analyzes the results in terms of service quality and user benefits. Finally, Section~\ref{sec:conclusion} concludes with final remarks and future research.
\section{Literature Review}
\label{sec:literature}

Dial-a-Ride services have been highlighted by several studies, particularly in the context of dispatching and routing optimization. \cite{cordeau2007darp} established several fundamental mixed-integer programming (MIP) formulations for the dial-a-ride problem (DARP).
These formulations address two distinct cases: one in which trip requests are known a priori and another in which trip requests arise dynamically in real time.
\cite{alonso-mora2017ondemand} were the first to address large-scale DARP, developing a heuristic that employs shareability graphs and cliques to generate feasible routes and a MIP model to select routes and perform matching of rider requests to vehicles. This approach was later enhanced by \cite{ZAC2018} through the use of zone paths.
\revision{Over time, numerous studies have sought to improve the scalability of solution methods for dial-a-ride, ridesharing, and ridesourcing problems under a variety of operational assumptions \citep{masoud2017decomposition, masoud2017real, regue2016car2work, lloret2017peer, tafreshian2020trip, zhang2021distributed}.}
\revision{The algorithm proposed by} \cite{riley2019column} was the first algorithm to serve all requests in a large-scale setting, using column generation and an exponentially increasing penalty function for unserved requests.
Hybrid methods have been proposed that combine routing recurrent requests offline while managing unexpected requests on-demand \citep{tafreshian2021proactive}.

As autonomous vehicle technology becomes more advanced and resilient, its  integration in transit and mobility services becomes increasingly more feasible. \citet{litman2020autonomous} discussed the impact of SAVs on transport planning, highlighting their potential benefits, including reduced traffic and parking congestion, improved mobility, enhanced safety, energy savings, and emission reductions. 
More recent studies have shown that SAVs can induce a modal shift toward shared mobility, and that fleets of SAVs operating in conjunction with public transit systems can yield greater benefits than when operating independently \citep{shaheen2016mobility, salazar2018interaction}.
\citet{levin2019linear} and \citet{shen2017embedding} proposed methods for including SAVs as last-mile connectors for transit networks. \citet{lam2016autonomous} designed a point-to-point public transit model serviced by SAVs. Simulation-based models have analyzed fleet sizing and operations for on-demand SAV fleets that serve first- and last-mile connections \citet{shen2018integrating,wang2018simulation}. In hybrid simulation-optimization approaches, 
\cite{pinto2020joint} designed a bi-level optimization framework where the upper level jointly determines transit route frequencies and SAV fleet sizes, and the lower level performs simulation for the traffic assignment of users. 
\cite{ng2024redesigning} extended this study and designed a multicommodity network flow model that incorporates zonal clustering, cost-aware transit design, and route generation.

\begin{figure}[!ht]
	\centering
	\includegraphics[width=0.7\linewidth]{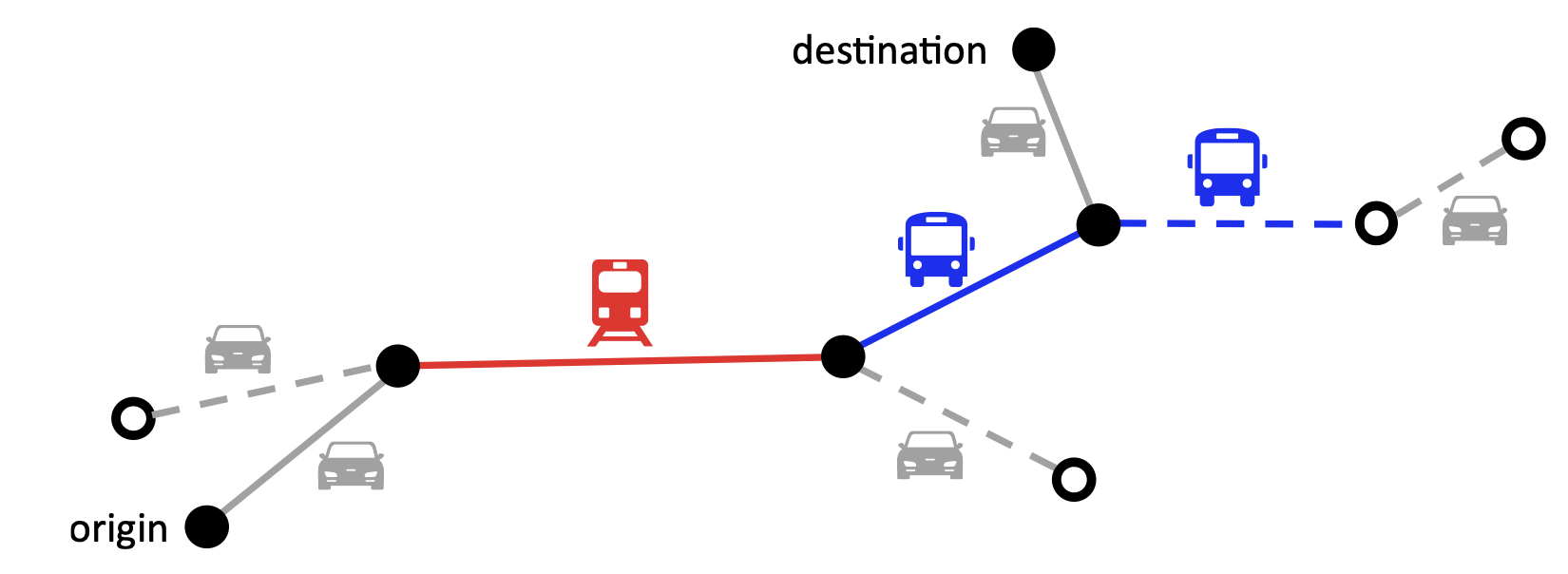}
	\caption{Example ODMTS with Passenger Path (solid lines)}
	\label{fig:odmts_example}
\end{figure}

MPC approaches have become increasingly important for repositioning in ridesharing and on-demand mobility systems, enabling improved routing efficiency, reduced waiting times, and lower operational costs. \cite{riley2020} developed an MPC that employs vector autoregression to forecast weekly demand time series patterns to optimize taxi relocations across zones. \cite{wei2023reinforcement} proposed a reinforcement learning-based approach to generate reposition recommendations for vehicles combined with a linear programming (LP) formulation to model driver compliance behavior. MPC \revision{has} become increasingly more relevant for autonomous mobility-on-demand (AMoD) systems. \cite{zhang2016model} introduced two MIP MPC formulations for vehicle rebalancing, where vehicle rebalancing decisions are made according to the demand levels of the waiting customers. \cite{tsao2018stochastic} extended this work to a stochastic MPC framework that incorporates predictions of short-term demand  uncertainty. \cite{zgraggen2019model} modeled AMoD services as feeders to a transit system and designed an MPC for AMoD repositioning using demand forecasting and LP. More recently, \cite{aalipour2024modeling} developed a finite-horizon control problem for MPC implementation in AMoD systems.

The case study presented in this paper is based on the MARTA Reach \revision{pilot}, which investigated the practical potential of ODMTS, an emerging type of microtransit system. The ODMTS system, initially proposed by \cite{maheo2019benders}, stems from the hub and shuttle transit system design introduced in earlier studies by \citet{campbell2005hub}. Figure~\ref{fig:odmts_example} illustrates an example of an ODMTS and the path of a single passenger.
In this example, a passenger is picked up by an on-demand shuttle for their first-mile leg near their origin and transported to a nearby train station.
The passenger is then instructed to take a train followed by a bus, both of which \revision{operate} on fixed schedules.
When the passenger arrives at the bus station closest to the destination, another on-demand shuttle serves the last-mile leg for the passenger.
This \href{https://sam.isye.gatech.edu/projects/demand-multimodal-transit-systems}{video} from the \citet{SAML2020-HowDemandMultimodal} demonstrates this process. 
ODMTS provides significant advantages, including greater accessibility to transit systems, enhanced mobility for those without personal vehicles, and a cost-efficient business model \citep{ kodransky2014connecting, lazarus2018shared, stiglic2018enhancing, mccoy2018integrating}.
Case studies in Canberra, Australia \citep{maheo2019benders}, Ann Arbor, Michigan \citep{ basciftci2023capturing}, and the metropolitan area of Atlanta, Georgia \citep{auad2021resiliency,lu2024impact, guan2023path, guan2025heuristic} demonstrate that ODMTS can simultaneously improve travel time, reduce system cost, and attract new passengers compared to existing transit systems. Furthermore, in addition to the aforementioned computational case studies, ODMTS has been piloted in Ann Arbor, Michigan \citep{van2019demand}, Atlanta, Georgia \citep{vanhentenryck2023reach}, and Savannah, Georgia \citep{Chatham}, and has been technologically enhanced throughout the pilots.

Other case studies have also evaluated transit systems that incorporate on-demand services. \cite{burstlein2021exploring} simulated the effects of varying DRT fleet sizes and vehicle capacities for first-mile transit services in Markham, Canada. In another study, \cite{sanaullah2021spatio} evaluated an eight-month pilot in the city of Belleville, Canada, where late night fixed-route transit was supplemented with on-demand services.

\subsection{Contributions}
Despite a rich body of literature on DARP, MPC, SAVs and ODMTS, several clear gaps remain. Existing MPC approaches for repositioning typically rely on extensive historical data, which poses two key limitations: \revision{1)} insufficient data is available during service startup, and \revision{2)} \revision{models are unable to} easily accommodate short-term fluctuations in demand. Prior studies on SAV integrations with transit systems do not provide comparisons between autonomous and non-autonomous services, particularly regarding key attributes such as dynamic vs non-dynamic re-routing and response times to requests when idling. 

This study addresses these shortcomings by leveraging enhanced dispatching models, an MPC that does not require extensive historical data, and scenario-based experiments that focus on the impacts of varying levels of autonomy for shuttles. Thus, this paper is the first study that models SAVs to address microtransit challenges associated with human driving behaviors and the first study to conduct a case study using ODMTS pilot data.
Specifically, the contributions of this paper are as follows:

\begin{enumerate}
    \item This paper presents \revision{IRDC}: a comprehensive framework that leverages SAVs to overcome the operational limitations of human-driven shuttles in microtransit systems and to model shuttle repositioning without relying on extensive historical data.
    \item To dispatch the shuttle fleet, this paper enhances the RTDARS dispatching model to incorporate human driver behavior modeling and introduces the RTDARS-SAV dispatching model to model the dynamic re-routing and immediate response times of SAV fleet operations. These two improvements lead to the design of scenario-driven experiments for microtransit systems, resulting in four distinct levels of fleet system competency and autonomy.
    \item This paper conducts a case study in the Atlanta \revision{metropolitan} area using data from the MARTA Reach pilot\revision{, which has not previously been used for experimental analysis in the academic literature. The case study evaluates} the effects of alternative fleet sizes, varying demand levels, and the integration of SAVs in microtransit systems. The results demonstrate the advantages of SAVs in microtransit systems in terms of both service quality and user convenience.
\end{enumerate}
\section{\revision{Integrated} Real-Time Dispatch and Control}
\label{sec:methodology}

\begin{figure}
    \centering
    \includegraphics[width=0.9\textwidth]{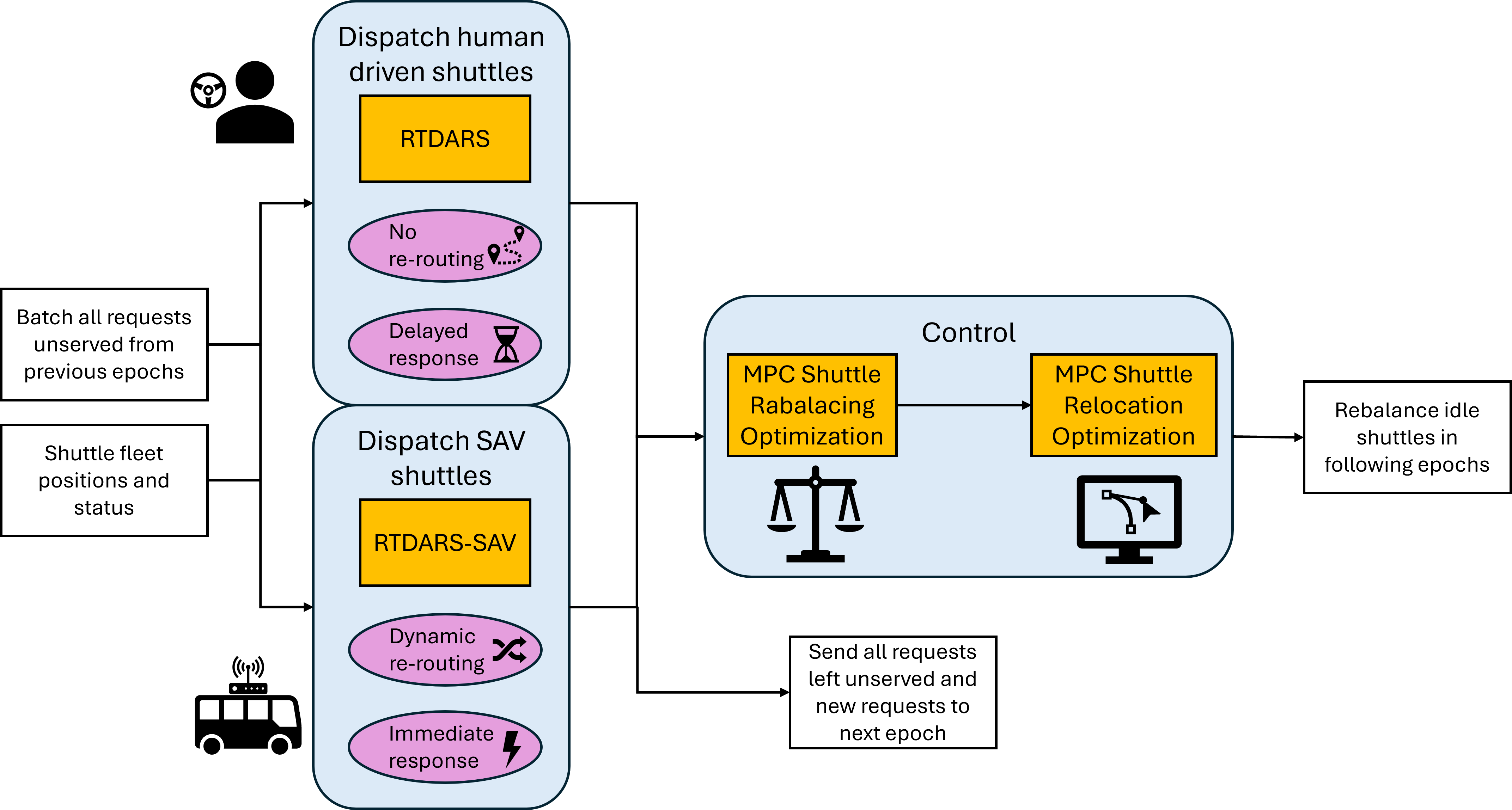}
    \caption{IRDC framework process for an epoch in a simulation setting}
    \label{fig:framework-pipeline}
\end{figure}

\revision{The Integrated Real-Time Dispatch and Control (IRDC) framework operates in an epoch-based setting, where shuttles are dispatched, routed, and rebalanced in response to dynamically arriving user requests, with the overarching goal of improving service quality by minimizing passenger waiting times. Figure~\ref{fig:framework-pipeline} illustrates the operational pipeline for a single epoch within the end-to-end simulation environment. The framework consists of two sequential components. The \emph{Dispatch} component assigns and routes shuttles to serve trip requests using either the \revision{Real-Time Dial-a-Ride System (RTDARS)} model for human-driven operations or the RTDARS-SAV model for SAV operations. The \emph{Control} component applies an MPC process to reposition idle shuttles toward areas of higher anticipated demand through two sequential optimization models: a shuttle rebalancing optimization model followed by a shuttle relocation optimization model.}

The RTDARS dispatching model extends from \cite{riley2019column} and models real-time human-driven shuttle system dynamics, processing online trip requests while managing the routing of shuttles across the network. \revision{Our extension addresses a limitation of \cite{riley2019column}, which assumed negligible response times for idling drivers. In practice, drivers may require non-negligible, and occasionally substantial, time to respond to trip assignments while idling, resulting in additional passenger waiting time. To capture this behavior, this study incorporates a sampling-based approach that models stochastic driver response delays to trip requests.}

\revision{The RTDARS-SAV dispatching model extends the RTDARS dispatching model to capture the operational capabilities of SAV fleets in a myopic setting, specifically zero-delay responses to trip requests and dynamic en route re-routing. SAVs can process incoming requests immediately upon arrival, enabling instantaneous assignment without delays caused by human driver response to request notifications. In addition, consistent with standard driver distraction policies adopted by transit agencies, human-driven microtransit operations prohibit shuttles from being re-routed once they are en route to their next stop because drivers would need to physically interact with the dispatching mobile application to acknowledge and respond to new trip requests, creating a potential source of distraction. SAVs are not subject to this restriction, as they can safely process trip requests while en route without compromising vehicle operation and can dynamically re-route in response to newly received requests. The RTDARS-SAV dispatching model incorporates these capabilities to more accurately represent SAV operations.}

\revision{The MPC framework rebalances idle shuttles to designated idle locations when they are not serving passengers in both the RTDARS and RTDARS-SAV dispatching models. The framework assumes that idle shuttles may only wait at designated idle locations rather than freely throughout the network, reflecting the traffic, safety, and land ownership constraints of microtransit operations. Given this operational constraint, the proposed MPC strategically repositions idle shuttles among the designated idle locations toward areas of higher anticipated demand.}

\revision{The models presented are epoch-specific but are applied to any epoch in the planning horizon.} The notation used throughout the paper is summarized in Table \ref{tab:notation}.

\subsection{Preliminaries}
\label{subsec:digital-twin}

Consider a finite time horizon $\mathcal{T}$, \revision{designated by each epoch $\tau \in [\mathcal{T}]$}. RTDARS processes requests by epochs, each of length $l$, and performs two tasks during each epoch:
1) batching the requests in the current epoch and 2) solving an optimization problem to assign and route unserved requests from previous epochs, minimizing the cost of service. \revision{It is important to note that requests arriving within the current epoch are not included in this optimization, as the full set of requests for the epoch is not known until the epoch has been fully realized.}

\begin{longtable}{p{0.2\textwidth}p{0.75\textwidth}}
\caption{Summary of notation.}
\label{tab:notation}\\
\toprule
\textbf{Notation} & \textbf{Description} \\
\midrule
\endfirsthead

\caption[]{Summary of notation (continued).}\\
\toprule
\textbf{Notation} & \textbf{Description} \\
\midrule
\endhead

\midrule
\multicolumn{2}{r}{Continued on next page} \\
\bottomrule
\endfoot

\bottomrule
\endlastfoot

\multicolumn{2}{l}{\textit{Indices and sets}}\\[0.25em]

\revision{$\mathcal{T}$} & Finite set of \revision{epochs} in the planning horizon.\\
$\tau$ & \revision{Epoch}, \revision{ $\tau \in [\mathcal{T}]$}.\\
$S$ & Set of virtual stop locations (pickup and dropoff locations).\\
$S_{\text{idle}} \subseteq S$ & Set of idle stops where shuttles are allowed to idle.\\
$V$ & Set of shuttles in the fleet.\\
$\revision{V_{\text{idle}}^\tau} \subseteq V$ & Set of idle shuttles with no current or future passengers \revision{at epoch $\tau$}.\\
$\revision{\hat V_{\text{idle}}^\tau \subseteq V_{\text{idle}}^\tau}$ & Set of idle shuttles that are not currently at an idle stop \revision{ at epoch $\tau$}.\\
\revision{$R^\tau$} & Set of all candidate shuttle routes at epoch $\tau$.\\
\revision{$R_v^\tau$} & Set of all candidate shuttle routes that can be served by shuttle $v \in V$ at epoch $\tau$.\\
\revision{$N$} & \revision{Set of all requests.}\\
\revision{$N^\tau$} & Set of unserved requests carried over from previous epochs \revision{at epoch $\tau$}.\\
\revision{$N_0^\tau$} & \revision{Set of all requests up to epoch $\tau$.}\\
\revision{$N_L^\tau \subseteq N_0^\tau$} & Set of recent requests in the consideration window $L$.\\
$D$ & Set of potential driver response times.\\
[0.5em]

\multicolumn{2}{l}{\textit{Request and route attributes}}\\[0.25em]

\revision{$p_n$} & Pickup location of request $n$, \revision{$p_n \in S$}.\\
\revision{$d_n$} & Dropoff location of request $n$, \revision{$d_n \in S$}.\\
\revision{$p_{n}^{\text{idle}}$} & \revision{Closest idle stop to $p_n$ in terms of travel time, $p_{n}^{\text{idle}} \in S_{\text{idle}}$.}\\
\revision{$e_n$} & Earliest possible pickup time of request $n$\revision{, $0 \leq e_n \leq \mathcal{T}l$.}\\
\revision{$a_{n}^r$} & Parameter equal to 1 if request $n$ is served by route $r$, and 0 otherwise.\\
$c_r$ & Total waiting time accumulated from all requests served by route $r$.\\[0.5em]

\multicolumn{2}{l}{\textit{Parameters}}\\[0.25em]

$l$ & Length of an epoch, unit in seconds.\\
$L$ & Length of the rolling consideration window for recent requests.\\
$Q_{\text{shuttle}}$ & Passenger capacity of each shuttle.\\
$g_n$ & Time dependent penalty for leaving request $n$ unserved in the current epoch.\\
$\delta$ & Incentivization parameter that scales the growth of penalties $g_n$.\\
$\revision{\gamma_s^\tau}$ & Number of recent requests mapped to idle stop $s$ (computed from \revision{$N_L^\tau$}).\\
$\rho^{v}_{s}$ & Travel time required to move shuttle $v$ to idle stop $s$.\\
$\revision{t_{\tau}}$ & Start time of epoch $\tau$.\\
$\hat d$ & Sampled driver response time, $\hat d \sim \text{Unif}(D)$.\\
$\Delta$ & Driver response time threshold.\\[0.5em]

\multicolumn{2}{l}{\textit{Decision variables}}\\[0.25em]

$\revision{y_r^\tau}$ & Binary variable equal to 1 if route $r \in \revision{R^\tau}$ is selected in the RTDARS master problem \revision{at epoch $\tau$}, and 0 otherwise.\\
$\revision{x_n^\tau}$ & Binary variable equal to 1 if request $n \in \revision{N^\tau}$ remains unserved \revision{at epoch $\tau$}, and 0 otherwise.\\
\revision{$\zeta_s^\tau$} & Integer variable for the number of shuttles to allocate to idle stop $s \in S_{\text{idle}}$ \revision{at epoch $\tau$}. \revision{The resulting value is subsequently used as a parameter in Model~\eqref{formulation:rvr-mip}}.\\
$\revision{z_s^\tau}$ & Auxiliary variable that acts as a proxy for $\revision{\zeta_s^\tau}$ in the MPC objective.\\
$\revision{w^{v,\tau}_{s}}$ & Binary variable equal to 1 if shuttle $v \in \revision{\hat V_{\text{idle}}^\tau}$ is relocated to idle stop $s \in S_{\text{idle}}$ \revision{at epoch $\tau$}, and 0 otherwise.\\[0.5em]

\multicolumn{2}{l}{\textit{Scenario descriptors}}\\[0.25em]

$\text{SFL $I$},\,\text{SFL $II$},$ & Shuttle functionality levels representing different driver or SAV capabilities and \\
$\text{SFL $III$},\,\text{SFL $IV$}$  & idle stop policies.\\

\end{longtable}

Let $S$ denote the set of virtual stop locations, which are potential user pickup and dropoff locations. Virtual stops are termed virtual as they do not require physical signage or infrastructure to mark their location, enabling more flexible placement within the service area. Shuttles are permitted to idle at idle stops, denoted by set \revision{$S_{\text{idle}} \subseteq S$}. Unlike virtual stops, idle stops are carefully selected to account for geographical constraints and property restrictions, and therefore cannot be freely located.

\revision{Let $N^\tau$ represent the set of requests at epoch $\tau$ that remain unserved from previous epochs. Each request is represented by a tuple $n = (p_n, d_n, p_{n}^{\text{idle}},e_n) \in N^\tau$, where $p_n \in S$ is the pickup location, $d_n \in S$ is the dropoff location, $p_{n}^{\text{idle}} \in S_{\text{idle}}$ is the closest idle stop to $p_n$ in terms of travel time, and $e_n$ is the earliest possible pickup time. For recent trips, the set $N_L^\tau$ denotes the collection of trip requests occurring within the consideration window $L$ from the current epoch $\tau$, which is updated at each epoch such that $N_L^\tau = \{ n \in N_0^\tau \mid \tau l - L \le e_n \le \tau l \}$. Maintaining only this set of recent requests allows the framework to adapt to evolving spatial demand patterns without relying on extensive historical data.}

\revision{Let $R^\tau$ denote the set of all routes in the study area \revision{ at epoch $\tau$. Each route $r \in R^\tau$ consists of a disjoint sets of requests served by a single shuttle, i.e., $r \subseteq N^\tau$}. These routes are iteratively generated through an optimization-based routine, the details of which can be found in \cite{riley2019column}. 
Let $c_r$ represent the total waiting time accumulated from all requests served by route $r \in R^\tau$.
The set $V$ denotes the set of shuttles, where each shuttle $v \in V$ could serve a subset of routes $\revision{R_v^\tau \subseteq R^\tau}$. The number of passengers per shuttle is limited by the shuttle capacity \revision{$Q_{\text{shuttle}}$}, which is constant across all shuttles. An idle shuttle, which does not have any current nor future planned passengers, is denoted as \revision{$v \in V_{\text{idle}}^\tau$}. Additionally, \revision{$\hat V_{\text{idle}}^\tau$} denotes the set of idle shuttles that are not currently at an idle stop.}

An unserved request $n \in \revision{N^\tau}$ incurs a time-dependent penalty \revision{$g_n = \delta2^{(\tau l - e_n)/10l}$}, which increases with each additional epoch until the request has been served. This prioritizes requests that have experienced longer waiting times. Here, $\delta$ is an incentivization parameter that weights scheduling in the earliest available epoch.
At the end of each epoch, new requests are introduced into the system and penalties for unserved requests from previous epochs are updated. RTDARS guarantees that all requests are ultimately served, leaving no passengers unserved after all epochs have been processed.

\begin{mini!}
%
{}
%
{\sum_{r \in \revision{R^\tau}} c_r \revision{y_r^\tau} + \sum_{n \in \revision{N^\tau}} g_n \revision{x_n^\tau},\label{eq:rtdars_master_obj}}
%
{\label{formulation:rtdars_master}}
%
{}
%
%
\addConstraint
{\sum_{r \in \revision{R^\tau}} \revision{y_r^\tau} a_n^r + \revision{x_n^\tau}}
{= 1 \quad \label{eq:rtdars_master_constr1}}
{\forall n \in \revision{N^\tau}}
\addConstraint
{\sum_{r \in \revision{R_v^\tau}} \revision{y_r^\tau}}
{= 1 \quad \label{eq:rtdars_master_constr2}}
{\forall v \in V}
\addConstraint
{\revision{x_n^\tau}}
{\in \{0,1\} \quad \label{eq:rtdars_master_constrNonneg1}}
{\forall n \in \revision{N^\tau}}
\addConstraint
{\revision{y_r^\tau}}
{\in \{0,1\} \quad \label{eq:rtdars_master_constrNonneg2}}
{\forall r \in \revision{R^\tau}}
\end{mini!}%

\subsection{RTDARS}

The epoch-based optimization considers both the waiting times incurred when serving requests and penalties associated with unserved requests in previous epochs.
\revision{Shuttles that are already committed to routes assigned in earlier epochs remain unavailable until they reach their next scheduled stop.}
To maintain service quality, routes for shuttles are constrained to ensure that passengers do not significantly deviate from their ideal paths.

\revision{Due to the combinatorial number of possible routes over the set of unserved trip requests, this study employs the column generation algorithm for RTDARS from \cite{riley2019column} to solve the optimization problem.}
The algorithm iteratively selects shuttle routes through a \revision{LP relaxation of the} master problem and generates new shuttle routes through a pricing subproblem \revision{using the dual values obtained from the relaxed master problem}, representing new columns in the master problem.
\revision{Upon reaching the stopping criterion, defined by a runtime limit equal to the epoch length $l$, the algorithm solves the master problem once more with integrality constraints enforced. \revision{It returns the optimal solution if optimality can be established over the generated columns within the runtime limit; otherwise, it returns the incumbent solution available at termination.} Although this approach may yield lower solution quality than a full branch-and-price algorithm, branch-and-price is generally too computationally expensive for rolling-horizon applications. Therefore, this computationally efficient approach is adopted to satisfy the real-time requirements of the framework.}

\revision{The master problem uses two binary decision variables: $\revision{y_r^\tau} = 1$ if route $r \in R^\tau$ is selected in the solution at epoch $\tau$, and $\revision{x_n^\tau} = 1$ if request $n \in N^\tau$ remains unserved by any selected route at epoch $\tau$.}
Model~\ref{formulation:rtdars_master} presents the master problem formulation for RTDARS column generation algorithm.
Objective~\eqref{eq:rtdars_master_obj} minimizes the total waiting times across selected routes and penalty costs from unserved requests.
Constraint~\eqref{eq:rtdars_master_constr1} ensures that requests are either served by a shuttle or left unserved \revision{at epoch $\tau$}. \revision{The auxiliary parameter $a_n^r$ is set to 1 if request $n$ is served by route $r$ and 0 otherwise.}
Constraint~\eqref{eq:rtdars_master_constr2} ensures that shuttles are assigned to a single route. \revision{Feasible compatibility between shuttles and requests is enforced through the pricing subproblem, which generates the set of feasible routes $R_v^\tau$ for each shuttle $v \in V$.}
Constraints~\eqref{eq:rtdars_master_constrNonneg1} and \eqref{eq:rtdars_master_constrNonneg2} define the binary conditions for decision variables $\revision{x_n^\tau}$ and $\revision{y_r^\tau}$, respectively.

\subsubsection{Modeling Driver Behavior in RTDARS}
\label{subsec:driver-behavior}
In practice, idling drivers exhibit inconsistent response times to trip assignments and rebalancing instructions.
Consider a full set of potential driver response times $D = \{d_1,\ldots,d_{|D|}\}$, where each response time is assumed to have an equal probability of occurrence. During the execution of RTDARS, if a shuttle is unoccupied during any epoch, the driver will act on the next instruction received with a uniformly sampled response time such that the observed response time $\hat{d} \sim \text{Unif}(D)$. Note that identical response times may appear within $D$, implying a higher probability of selecting response times that occur more frequently in the set. Although drivers may not respond at all in practice, requiring requests to be reallocated, we assume that all drivers respond within a threshold $\Delta$ such that any observed response time satisfies $\hat{d} \leq \Delta$.

\subsection{RTDARS-SAV}
\label{subsec:avs}

\revision{In RTDARS for human-driven shuttles, any routing decision made at epoch $\tau$ assumes that a shuttle cannot be reassigned until it completes travel to its next scheduled stop, reflecting the operational limitations associated with human-driven services. In contrast, RTDARS-SAV removes this restriction by allowing SAVs to be re-routed while traveling between stops.}

\revision{When optimizing at epoch $\tau$ with start time $t_{\tau}$, the simulator identifies SAVs that are currently en route and estimates their geographic locations at the beginning of the next decision epoch, $t_{\tau+1}$. These locations are obtained by interpolating each vehicle's position along the road-network polyline connecting its previous and next scheduled stops based on the elapsed travel time. Rather than treating the next scheduled stop as the routing origin, the optimization routine uses this estimated mid-segment location as the starting point for subsequent routing decisions. Consequently, newly received trip requests can be incorporated immediately without requiring the SAV to first reach its previously assigned stop, enabling dynamic en route re-routing and improving operational responsiveness.}

Furthermore, there is no need to model driver behavior, as SAVs exhibit zero-delay responses to trip requests while idling. In RTDARS-SAV, the driver behavior modeling component described in Section~\ref{subsec:driver-behavior} is therefore omitted.

\subsection{Model Predictive Control}
\label{subsec:mpc}

This study introduces a procedure to rebalance inactive shuttles to designated idle stops that have potential for higher user demand in the nearby area \revision{without depending on extensive data availability, which may not be available in many deployment contexts at the early stages.}

The objective is to rebalance idle shuttles toward areas of high demand while ensuring that shuttles only idle at stops within \revision{$S_{\text{idle}}$}. Equation~\eqref{eq:c_s_equation} computes the number of recent requests $\revision{\gamma_s^\tau}$ mapped to each idle stop \revision{at epoch $\tau$}, where $\mathbb{I}$ is a binary indicator function. \revision{This leverages the encoding of $p_n^{\text{idle}}$ to associate requests with their nearest idle stop, thereby directing shuttles toward idle stops in areas of higher demand.}

\begin{equation}
    \revision{\gamma_s^\tau} = \sum_{n \in \revision{N_L^\tau}} \mathbb{I}[\revision{p_n^{\text{idle}}} = s] \quad \forall s \in \revision{S_{\text{idle}}}
    \label{eq:c_s_equation}
\end{equation}

Model~\eqref{formulation:vr-ratio} rebalances the shuttle fleet, determining the desired number of shuttles, defined as the decision variable $\revision{\zeta_s^\tau}$, to allocate at each idle stop \revision{$s \in S_{\text{idle}}$ at epoch $\tau$}, given the values of $\revision{\gamma_s^\tau}$. Objective~\eqref{eq:vr-ratio-obj} minimizes the total ratio of requests, $\revision{\gamma_s^\tau}$, to shuttles, $\revision{z_s^\tau}$, a proxy for $\revision{\zeta_s^\tau}$, across all idle stops. Constraint~\eqref{eq:vr-ratio_constr2} preserves the validity of the objective function by preventing division errors via $\revision{z_s^\tau}$ when $\revision{\zeta_s^\tau} = 0$ and implicitly doubles the penalty associated with assigning zero shuttles to an idle stop compared to assigning one. Constraint~\eqref{eq:vr-ratio_constr1} ensures that the total number of shuttles assigned across all idle stops equals the total number of idle shuttles available. The model considers only the set \revision{$\hat{V}_{\text{idle}}^\tau$} rather than the \revision{$V_{\text{idle}}^\tau$} to avoid relocating shuttles already at an idle stop, which incurs higher cost of resources. Constraint~\eqref{eq:vr-ratio_naturalconstr} defines $\revision{\zeta_s^\tau}$ as a non-negative integer decision variable. \revision{Lastly, constraint~\eqref{eq:z_naturalconstr} defines $\revision{z_s^\tau}$ as a positive integer decision variable while permitting the value $0.5$.}

\begin{mini!}
    %
    {}
    %
    {\sum_{s \in \revision{S_{\text{idle}}}} \frac{\revision{\gamma_s^\tau}}{\revision{z_s^\tau}},\label{eq:vr-ratio-obj}}
    %
    {\label{formulation:vr-ratio}}
    %
    {}
    %
    %
    \addConstraint
    {\revision{z_s^\tau}}
    {= \max\{0.5,\revision{\zeta_s^\tau}\} \quad \label{eq:vr-ratio_constr2}}
    {\forall s \in \revision{S_{\text{idle}}}}
    \addConstraint
    {\sum_{s \in \revision{S_{\text{idle}}}} \revision{\zeta_s^\tau}}
    {= |\revision{\hat{V}_{\text{idle}}^\tau}| \quad \label{eq:vr-ratio_constr1}}
    {}
    \addConstraint
    {\revision{\zeta_s^\tau}}
    {\in \mathbb{N}^0 \quad \label{eq:vr-ratio_naturalconstr}}
    {\forall s \in \revision{S_{\text{idle}}}}
    \addConstraint
    {\revision{\revision{z_s^\tau}}}
    {\in \revision{\mathbb{N}^+ \cup \{0.5\}} \quad \label{eq:z_naturalconstr}}
    {\revision{\forall s \in S_{\text{idle}}}}
\end{mini!}%

Model~\eqref{formulation:vr-ratio} returns the number of shuttles $\revision{\zeta_s^\tau}$ desired at idle stop $s \in \revision{S_{\text{idle}}}$ \revision{at epoch $\tau$}. However, it does not identify which specific shuttles should be relocated. To address this, Model~\eqref{formulation:rvr-mip} presents a MIP for shuttle relocation. \revision{Model~\eqref{formulation:rvr-mip} uses the optimal values of $\zeta_s^\tau$ obtained from Model~\eqref{formulation:vr-ratio} as fixed parameters, thereby treating them no longer as decision variables but as given inputs for the relocation problem.} The binary decision variable $\revision{w_s^{v,\tau}} = 1$ if shuttle $v \in \revision{\hat{V}_{\text{idle}}^\tau}$ is chosen to relocate to stop $s \in \revision{S_{\text{idle}}}$ at epoch $\tau$. Objective~\eqref{eq:rvr-mip-obj} minimizes the total travel time $\rho_{s}^v$ to move each shuttle $v \in \revision{\hat{V}_{\text{idle}}^\tau}$ to each idle stop $s \in \revision{S_{\text{idle}}}$. \revision{This travel time is not restricted to a single epoch $\tau$ and may span multiple epochs until the shuttle reaches its assigned idle stop.} Constraint~\eqref{eq:rvr-mip_constr1} ensures the correct number of shuttles are assigned to each idle stop. Constraint~\eqref{eq:rvr-mip_constr2} ensures that all shuttles relocate to exactly one idle stop. Lastly, constraint~\eqref{eq:rvr-mip_binaryconstr} defines binary conditions for $\revision{w_s^{v,\tau}}$.

It is important to note that Model~\eqref{formulation:vr-ratio} could induce tied solutions under Objective~\eqref{eq:vr-ratio-obj}, which can influence the solution of Model~\eqref{formulation:rvr-mip}. Appendix~\ref{appendix:rr-opt-w-tie-breaking} extends the model to incorporate tie-breaking procedures that resolve such cases. \revision{Both Models~\eqref{formulation:vr-ratio} and~\eqref{formulation:rvr-mip} are solved directly using a commercial optimization solver without requiring specialized large-scale solution algorithms, as both models can be solved to optimality with negligible computational effort.}

\begin{mini!}
    %
    {}
    %
    {\sum_{v \in \revision{\hat{V}_{\text{idle}}^\tau}} \sum_{s \in \revision{S_{\text{idle}}}} \rho_{s}^v \revision{w_s^{v,\tau}},\label{eq:rvr-mip-obj}}
    %
    {\label{formulation:rvr-mip}}
    %
    {}
    %
    %
    \addConstraint
    {\sum_{v \in \revision{\hat{V}_{\text{idle}}^\tau}} \revision{w_s^{v,\tau}}}
    {= \revision{\zeta_s^\tau} \quad \label{eq:rvr-mip_constr1}}
    {\forall s \in \revision{S_{\text{idle}}}}
    \addConstraint
    {\sum_{s \in \revision{S_{\text{idle}}}} \revision{w_s^{v,\tau}}}
    {= 1 \quad \label{eq:rvr-mip_constr2}}
    {\forall v \in \revision{\hat{V}_{\text{idle}}^\tau}}
    \addConstraint
    {\revision{w_s^{v,\tau}}}
    {\in \{0,1\} \quad \label{eq:rvr-mip_binaryconstr}}
    {\forall s \in \revision{S_{\text{idle}}}, v \in \revision{\hat{V}_{\text{idle}}^\tau}}
\end{mini!}%

\section{Case Study}
\label{sec:counterfactual-experiments}

This study uses data from the MARTA Reach \revision{pilot} in Atlanta by \cite{vanhentenryck2023reach}. The MARTA Reach pilot integrated on-demand shuttles to serve users with MARTA’s fixed-route bus and rail services, representing a pivotal development towards ODMTS in practice. This integration operated in conjunction with the broader transit network and in close cooperation with the MARTA. \revision{Importantly, the operational framework employed in the MARTA Reach \revision{pilot} closely aligns with the IRDC framework proposed in this study, where human-driven shuttles are dispatched, routed, and controlled in response to dynamically arriving user requests. Therefore, the operational data collected from the pilot provides a meaningful real-world case study for evaluating the proposed framework.}

\subsection{Pilot Operations}
\label{subsec:shuttle-operations}

The MARTA Reach pilot operated for six months, from March 1st to August 31st, 2022.
Daily operations went from 6 A.M. to 7 P.M. on weekdays, including federal holidays, for a total of 132 service days.
Passengers within the pilot zones could request trips by providing their origins and destinations through a mobile application or via telephone.
Operations were divided into two phases.
Phase 1 lasted from the pilot's introduction on March 1st, 2022 to May 15, 2022, where MARTA Reach operated in three distinct zones: Belvedere, West Atlanta, and Gillem.
Phase 2 lasted from May 16, 2022 to the pilot's end on August 31, 2022.
During this phase, the three initial zones all expanded, and North Fulton was included as an additional zone. All requests were geo-fenced such that trip pickup and dropoffs only occurred within the same zone. Trips can either connect users to a transit station or serve as direct trips, where users travel directly from their pickup to \revision{dropoff} locations. Users are charged a fixed fare regardless of trip type. In particular, passengers connecting to a fixed-route service pay the same fare and are not required to pay an additional fee for the transfer. Full details of the MARTA Reach pilot are available in \citet{vanhentenryck2023reach}.

This study focuses on 2 of the 4 MARTA Reach zones: West Atlanta and Belvedere.
These two zones accounted for the majority of ridership, with 58.45\% and 34.81\% of all requests being served from West Atlanta and Belvedere, respectively.
Furthermore, both zones included portions of the MARTA rail network, enabling the integration of on-demand shuttle services with fixed-route transit and supporting multimodal connectivity. In Phase 2, the West Atlanta zone operated with six shuttles during the pilot and the Belvedere zone operated with four.

For experimental analysis, this study uses MARTA Reach ridership data from August 31, 2022, the day with the highest ridership observed during the pilot.

\begin{figure}[!ht]
    \begin{subfigure}[h]{\textwidth}
        \centering
        \includegraphics[width=.9\textwidth]{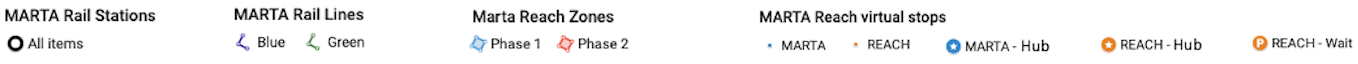}
    \end{subfigure}
    \centering
    \begin{subfigure}[h]{0.45\textwidth}
        \centering
        \includegraphics[width=\textwidth]{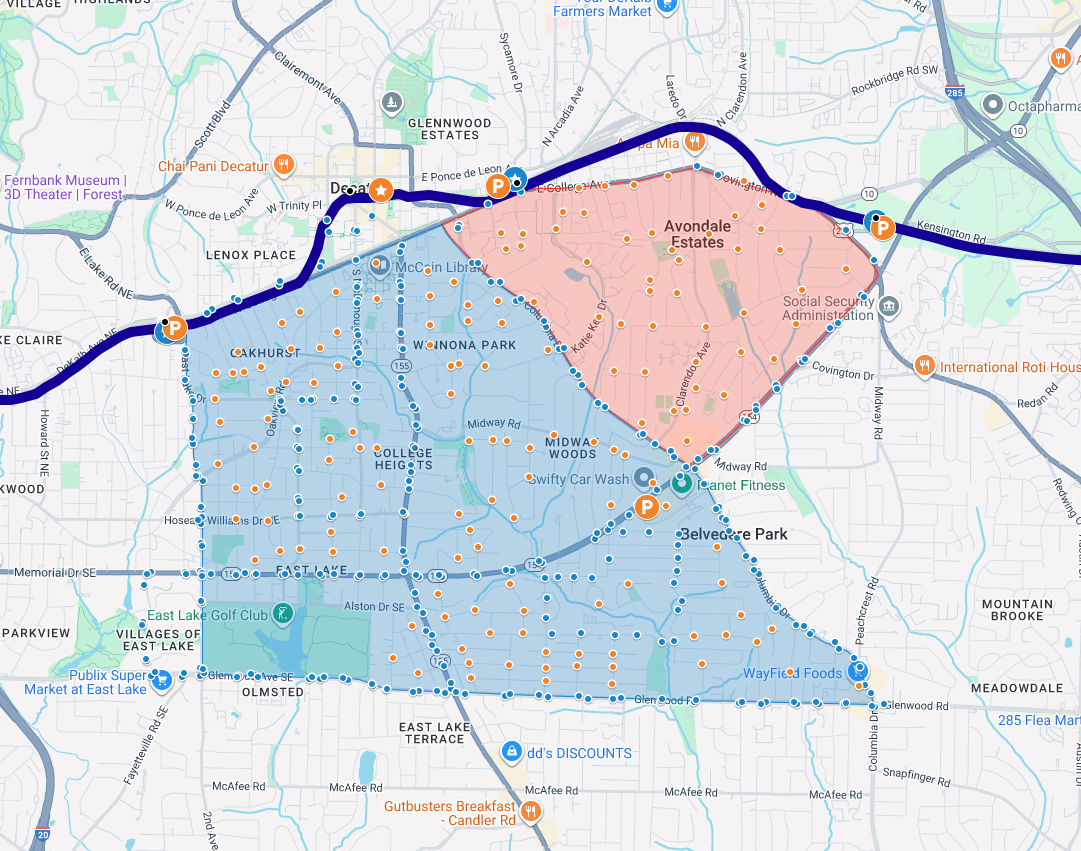}
        \caption{Belvedere}
        \label{fig:Belvedere_virtual_stops}
    \end{subfigure}
    \begin{subfigure}[h]{0.45\textwidth}
        \centering
        \includegraphics[width=\textwidth]{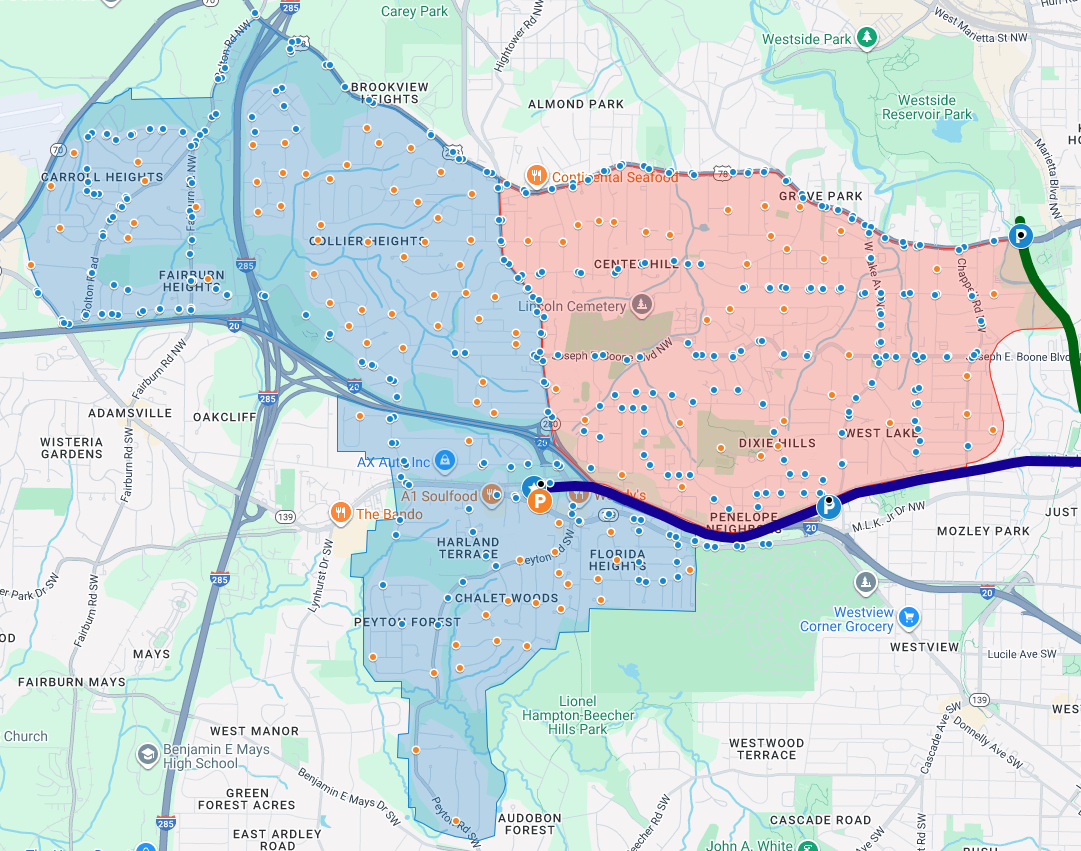}
        \caption{West Atlanta}
        \label{fig:WestAtlanta_virtual_stops}
    \end{subfigure}
\caption{Belvedere and West Atlanta \revision{MARTA} Reach zones (blue zones correspond to phase 1 deployment areas and pink zones correspond to phase 2 expansions of deployment areas)}
\label{fig:virtual_stops_zones}
\end{figure}

Shuttle operations were conducted with eight-passenger vehicles across two daily driver shifts: 6 A.M. to 1 P.M. and 1 P.M. to 7 P.M.
When receiving a trip request, the pilot system's optimization routine determined vehicle assignments. Safety \revision{protocols} of MARTA dictated that drivers were not permitted to handle on-demand received requests or make routing changes while en route between stops.
Shuttle pickup and dropoff operations were restricted to virtual stops, designated via geosensing technology implemented during the pilot.
Virtual stops were pre-approved by MARTA for both pedestrian and driver safety.
Unlike conventional bus stops, virtual stops did not necessarily have physical signage.
Virtual stops were positioned to sufficiently cover the MARTA Reach zones, providing convenient passenger access while maintaining operational efficiency and safety.

\begin{table*}[!htb]
    \centering
    \begin{tabular}{l r r r r r}
    \toprule 
    Zone & Total ($|S|$)  & \makecell{Existing MARTA \\ stops/stations} & \makecell{New MARTA \\ Reach stops} & \makecell{Idle stops \\ ($|S_{\text{idle}}|$)} & \makecell{\# Shuttles \\ ($|V|$)}\\
	\midrule
    Belvedere & 	465 & 	308 & 	157 & 	4 & 	 4\\ 
	West Atlanta & 	485 & 	374 & 	111 & 	3 & 	 6\\ 
    \bottomrule
    \end{tabular}
\caption{Zone Details of Pilot on August 31, 2022}
\label{tab:virtual_stops}
\end{table*}

Figure~\ref{fig:Belvedere_virtual_stops} and~\ref{fig:WestAtlanta_virtual_stops} display the virtual stops and idle locations for the Belvedere and West Atlanta zones, respectively.
The blue markers represent MARTA fixed-route stops and were essential for multimodal connections to bus and rail lines.
The orange markers represent additional stops introduced during MARTA Reach to serve passengers.
When not serving passengers, shuttles were restricted to idling in designated idle locations selected by MARTA, which prevents unnecessary congestion, safety hazards, and land ownership conflicts.
Drivers were required to proceed to one of the approved idle locations in their zone when not assigned to any active requests. Table~\ref{tab:virtual_stops} summarizes the makeup of virtual and idle stops, as well as the number of operational shuttles, in Belvedere and West Atlanta.

\subsection{Experimental Scenarios}
\label{subsec:experimental-scenarios}

The experimental scenarios in this study are defined by three attributes: fleet size, fleet functionality, and ridership level.
The distinct combinations of all these attributes constitute the total set of experimental scenarios simulated for each zone.
Table~\ref{tab:scenarios} summarizes the scenario attributes for Belvedere and West Atlanta.

\begin{table}[!ht]
\centering
\begin{tabular}{lcl}
\hline
Zone        & \multicolumn{1}{l}{Belvedere} & West Atlanta              \\ \hline
Fleet Sizes ($|V|$) & 3,4,5                         & \multicolumn{1}{c}{5,6,7} \\
SFLs       & \multicolumn{2}{c}{$I$, $II$, $III$, $IV$}              \\
Ridership Levels ($|N|$)   & \multicolumn{2}{c}{Base (1x), 2x, 3x}                          \\ \hline
\end{tabular}
\caption{Summary of Scenario Makeups}
\label{tab:scenarios}
\end{table}

\subsubsection{Fleets} Three fleet sizes are examined for each zone.
The baseline fleet size corresponds to that used during the MARTA Reach \revision{pilot}: four shuttles in Belvedere and six shuttles in West Atlanta.
The other two fleet size configurations for each zone are created by incrementing or decrementing the baseline fleet size by one vehicle.

\subsubsection{Shuttle Functionality Levels} The shuttle fleets are classified into four shuttle functionality levels (SFLs), reflecting the degree of automation and routing capability, summarized below:
\begin{itemize}
    \item SFL $I$ - Level 1 functionality: Human-driven shuttles with variable response times to passenger requests while idling at stops. This serves as the base SFL for modern microtransit pilot systems.
    \item SFL $II$ - Level 2 functionality: Human-driven shuttles with zero-delay response times when idling at stops
    \item SFL $III$ - Level 3 functionality: SAVs with zero-delay response times and dynamic re-routing capabilities to accommodate on-demand requests. While SAVs do not require any labor breaks, they do require maintenance such as recharging and thus continue to operate in shifts.
    \item SFL $IV$ - Level 4 functionality: SFL $III$ but with idle stops relocated to the most frequent origin points of trip requests at the start of each shift.
\end{itemize}

Table~\ref{tab:summary-sfls} provides a concise summary of the attributes of each SFL and the dispatching models used to solve them. SFL $I$ represents operational attributes of the MARTA Reach pilot, while SFL $III$ reflects those of SAV operations. SFLs $II$ and $IV$ serve as idealized operational performance scenarios corresponding to SFLs $I$ and $III$, respectively.

\begin{table}[!ht]
\centering
\begin{tabular}{l| l l cc}
\toprule
\revision{SFLs}        & Dispatching Model & Driver & Delay & Idle Locations            \\ \midrule 
$I$ & RTDARS & Human & Uniform & Predetermined and fixed    \\
$II$        & RTDARS  & Human & 0 & Predetermined and fixed         \\
$III$    & RTDARS-SAV & SAV  & 0  & Predetermined and fixed                     \\ 
$IV$    & RTDARS-SAV  & SAV  & 0  & Unrestricted and flexible                    \\ \bottomrule
\end{tabular}
\caption{Summary of SFL Attributes}
\label{tab:summary-sfls}
\end{table}

The idle relocation procedure for SFL $IV$ occurs twice daily and assumes full knowledge of all requests throughout the day. At the start of each shift, idle stops are repositioned based on the most common trip origins observed during the respective shift.
The number of idle stops remains constant across all experiments and only locations are updated.

\subsubsection{Ridership Sampling} 
To evaluate the impact of higher ridership levels on system performance, this study scales ridership by sampling trip requests from the days preceding August 31, 2022, the base experimental day, until a desired ridership level is achieved.
This allows consideration of double and triple the baseline trip volumes for both Belvedere and West Atlanta zones.
Only requests within Phase 2 of MARTA Reach are included as the base experimental day occurred during Phase 2 operations.
Trips from a user already sampled were included only if its time window does not overlap with any existing time window of the user. This ensures that synthetic ridership increases reflect realistic temporal distributions without duplicating passenger activity.

\subsubsection{Pilot Driver Behavior Sampling}
\label{sssec:pilot-driver-sampling}
During the MARTA Reach pilot, observed driver response delays for trip requests introduced challenges that adversely affected operations. \revision{Although the MARTA Reach backend software automatically issued dispatch alerts to drivers whenever a new trip request was received, drivers were still responsible for acknowledging and responding to these requests. Each request triggered both a pop-up notification and an audible alert, ensuring that drivers were notified immediately. Nevertheless, delays in driver responses still occurred despite the automated notification system.} As such, an analysis of driver response times to passenger trip requests was conducted, and the resulting response time distribution was incorporated into the experimental design.

To ensure reliability, the analysis focused on the month of August, which exhibited the highest ridership, stable service operations, and full deployment of all MARTA Reach functionalities \citep{vanhentenryck2023reach}. 1619 trips were analyzed in which shuttles were idle and subsequently dispatched to a passenger’s requested origin stop. This subset was selected because delays in such cases could be more reasonably attributed to driver behavior, whereas delays during active service were typically operational in nature and not indicative of driver inattention.

\begin{figure}[h]
    \centering
    \includegraphics[width=.8\textwidth]{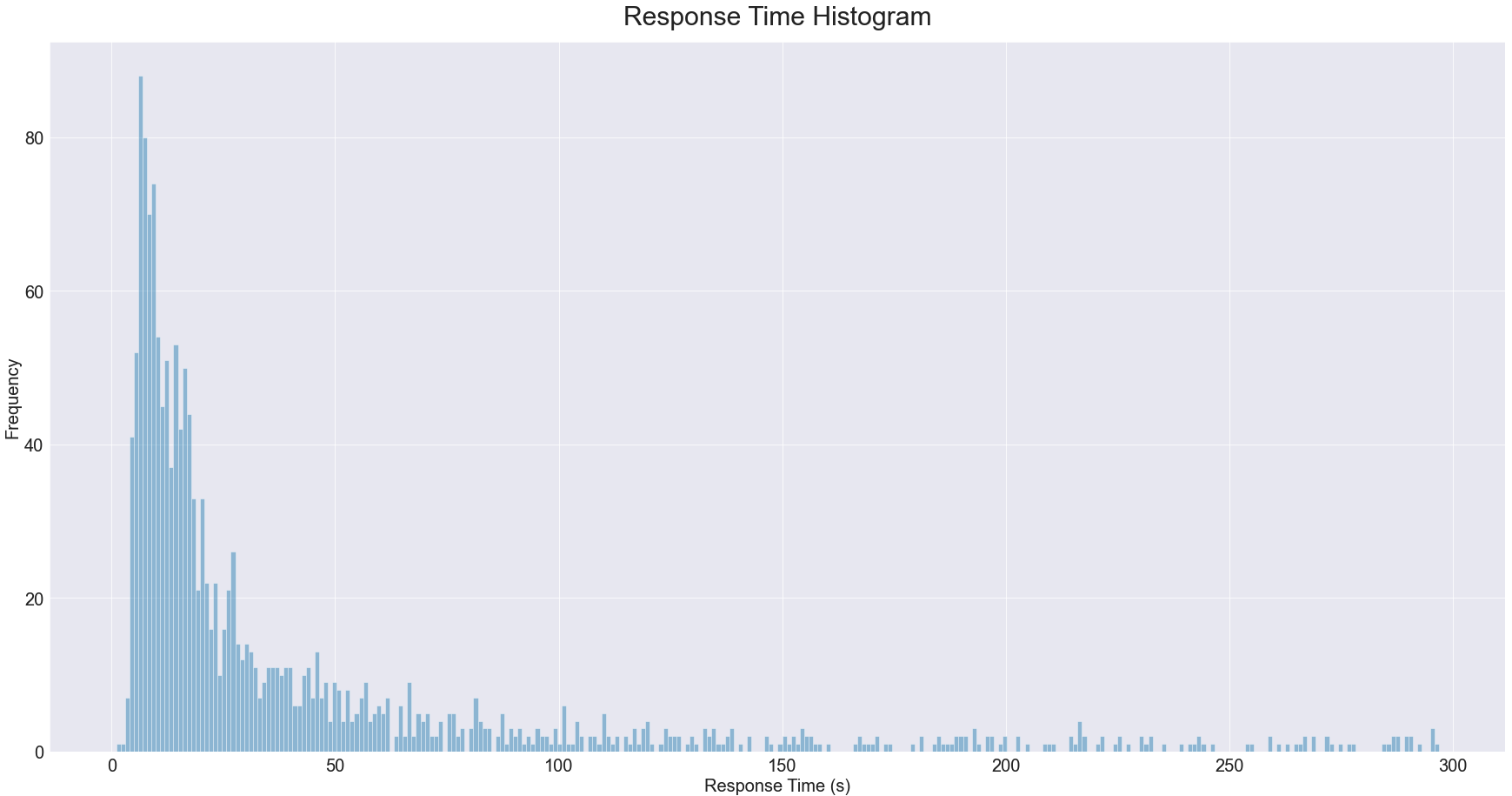}
    \caption{Driver Response Time Distribution}
\label{fig:response_time_distribution}
\end{figure}

Figure~\ref{fig:response_time_distribution} illustrates the distribution of driver response times, which is right-skewed with a mean response time of 43.14 seconds and a median response time of 18 seconds. The distribution indicates that \revision{the majority of requests received a prompt driver response, typically} within a minute. However, the extended right tail reveals occasional prolonged response times, suggesting intermittent issues related to driver attentiveness or delayed responsiveness. \revision{Although most observed response delays were relatively short, resulting in only modest differences between SFLs $I$ and $II$ on average, the extended right tail indicates that occasional long response delays can occur and substantially affect operations. In some extreme instances, the driver response delay may even exceed the shuttle travel time required to reach the passenger after accepting the request, increasing not only overall response times but also the variability and perceived reliability of the service from the passenger's perspective.}

\subsection{Experimental Setup}
\label{subsec:experimental-setup}

\begin{table}[!t]
	\centering
	\begin{tabular}{ll}
		\toprule
		Parameter & Value \\
		\midrule
		Lower Bound Time & 6 (MARTA Reach operations begin at 6 A.M.) \\
		Upper Bound Time & 19 (MARTA Reach operations end at 7 P.M.) \\
        $Q_{\text{shuttle}}$ & 6 (minivan DRT microtransit services) \\
        $l$ & 30 seconds (\cite{riley2019column}) \\
        $\delta$ & 420 seconds (\cite{riley2019column}) \\
        $L$ & 1 hour (MARTA Reach) \\
        $\Delta$ & 300 seconds (MARTA Reach)\\
		\bottomrule
	\end{tabular}
	\caption{Relevant Experimental Parameters}
	\label{tab:sim-paramters}
\end{table}
With all combinations of fleet sizes, modes, and ridership levels, a total of 36 experimental scenarios were simulated for each zone, with 72 total scenarios.

The time and distance between virtual stops was estimated using real data obtained during the pilot period and data from OpenStreetMap using the Graphhopper API \citep{OpenStreetMap}. Baseline values were initially obtained from OpenStreetMap, which provided the shortest possible time and distance values. These values were subsequently scaled to reflect congested traffic using travel time collected from the pilot zones. 
Outliers resulting from glitches or systematic errors in the data were excluded, and a simple linear regression was fitted using the remaining data. 

For all scenarios, shuttles were initially distributed uniformly across idle stops within each zone. For SFL $I$, driver response times $D$ were sampled from the response times of MARTA Reach drivers discussed in Sections~\ref{subsec:driver-behavior} and~\ref{sssec:pilot-driver-sampling}.

The base level of ridership for the Belvedere zone considers 39 completed trips (43 users) for the base day of August 31, 2022.
With sampling from preceding pilot days, 2x ridership considers 78 completed trips (82 users), and 3x ridership considers 117 completed trips (122 users).
West Atlanta considers 82 completed trips (89 users) from the base day, 164 completed trips (\revision{182} users) for 2x ridership, and 246 completed trips (\revision{270} users) for 3x ridership. Each trip could have multiple users, accommodating up to four for a single trip request.

\subsection{\revision{Simulation Environment}}
\label{subsec:sim-env}
\revision{All proposed experiments were conducted within an end-to-end simulation environment embedded into a simulator. The simulator models the dynamic assignment, routing, and rebalancing of shuttles within the study area in response to online trip requests by incorporating all components of the IRDC framework. It operates in an online setting, where trip requests are revealed dynamically and no future request information is available. The RTDARS column generation algorithm was originally developed by \cite{riley2019column} to efficiently solve large-scale instances of trip demand and transportation networks, including a case study based on New York City. Since the MARTA Reach case study considered in this paper is substantially smaller, the algorithm can easily solve the dispatching problem. To further improve computational performance, the simulator employs multi-threading techniques to parallelize computationally intensive tasks. The simulator was implemented in C++ 17 and Models~\eqref{formulation:rtdars_master},~\eqref{formulation:vr-ratio}, and~\eqref{formulation:rvr-mip} were solved using commercial solver Gurobi 9.5.0. All computational experiments were performed on a Linux high-performance computing cluster managed by HTCondor, using default HTCondor CPU resource assignments. Jobs were dispatched to available worker nodes equipped with AMD EPYC Rome or Genoa processors or Intel Xeon Haswell, Skylake, or Cascade Lake processors.}

Table~\ref{tab:sim-paramters} summarizes the key \revision{simulation} parameters. The lower and upper bound times match the duration for a single day of MARTA Reach. $l$ and $\delta$ were set to 30 seconds and 420 seconds, respectively, consistent with the effective parameter values identified by \cite{riley2019column} when balancing computational efficiency and request frequency. The trip consideration window $L$ was set to one hour and the response time threshold $\Delta$ was set to 300 seconds, identical to the MARTA Reach pilot. Shuttle capacity was set to 6, consistent with a typical minivan. Minivans were chosen over sedans to account for the larger vehicle capacities often required in public transit operations. Although sedans may be more cost-efficient, transit agencies are non-profit, where service quality and accessibility are prioritized. Furthermore, this configuration supports possible integration with paratransit services. First, paratransit fleets typically accommodate six or more passengers, allowing these services to leverage existing vehicles rather than investing in new shuttles. Second, \revision{besides MARTA Reach}, ODMTS was \revision{once} piloted alongside paratransit services, and in the future paratransit services can transition into on-demand operations using a reservation-based system \citep{Chatham}.

\begin{figure}[!ht]
    \centering
    \begin{subfigure}[]{0.45\textwidth}
        \includegraphics[width=\textwidth]{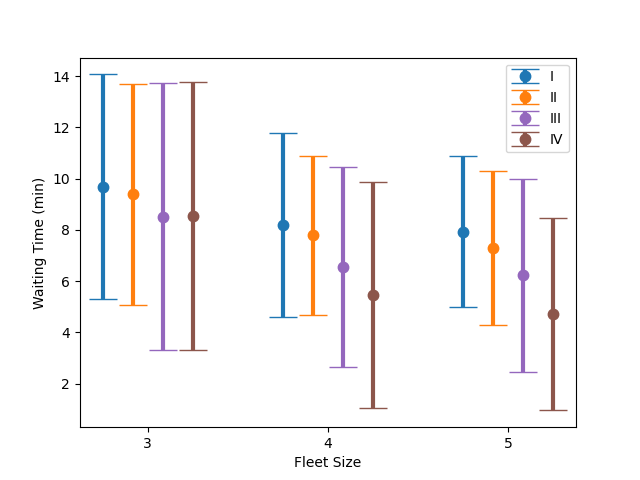}
        \caption{Belvedere}
        \label{fig:wait-belvedere-mode-base}
    \end{subfigure}
    \begin{subfigure}[]{0.45\textwidth}
        \includegraphics[width=\textwidth]{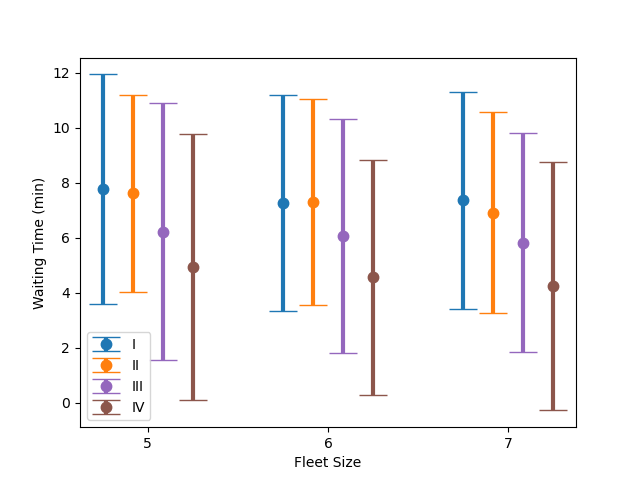}
        \caption{West Atlanta}
        \label{fig:wait-westatlanta-mode-base}
    \end{subfigure}
\caption{Average Waiting Time Per Trip Across SFLs at Base Ridership Level}
\label{fig:avg-wait-time-per-trip-mode-base}
\end{figure}

\begin{figure}[!ht]
    \centering
    \begin{subfigure}[]{0.45\textwidth}
        \includegraphics[width=\textwidth]{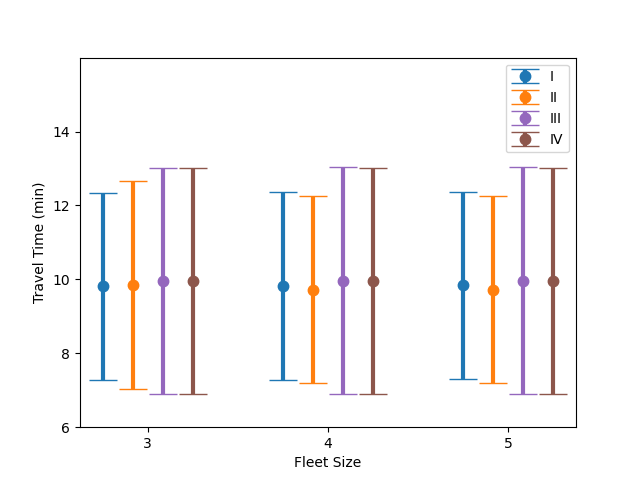}
        \caption{Belvedere}
        \label{fig:travel-belvedere-mode-base}
    \end{subfigure}
    \begin{subfigure}[]{0.45\textwidth}
        \includegraphics[width=\textwidth]{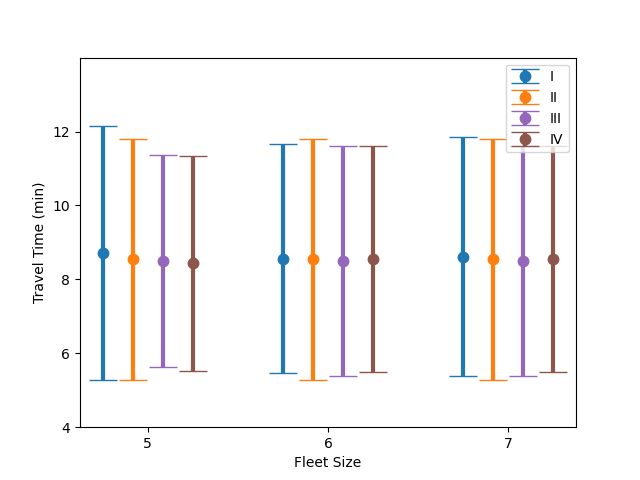}
        \caption{West Atlanta}
        \label{fig:travel-westatlanta-mode-base}
    \end{subfigure}
\caption{Average Travel Time Per Trip Across SFLs at Base Ridership Level}
\label{fig:avg-travel-time-per-trip-mode-base}
\end{figure}

\begin{figure}[!ht]
    \centering
    \begin{subfigure}[]{0.45\textwidth}
        \includegraphics[width=\textwidth]{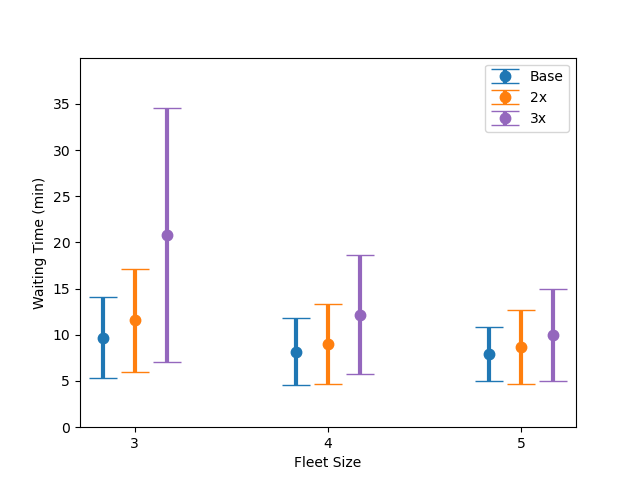}
        \caption{Belvedere - SFL $I$}
        \label{fig:wait-westatlanta-rider-db}
    \end{subfigure}
    \begin{subfigure}[]{0.45\textwidth}
        \includegraphics[width=\textwidth]{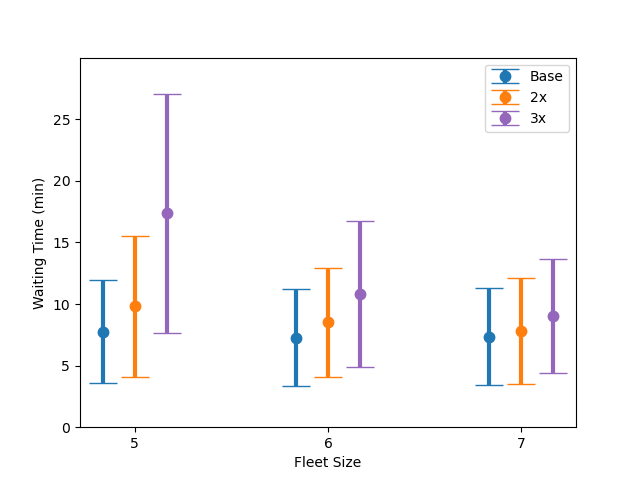}
        \caption{West Atlanta - SFL $I$}
        \label{fig:wait-belvedere-rider-db}
    \end{subfigure}
    \begin{subfigure}[]{0.45\textwidth}
        \includegraphics[width=\textwidth]{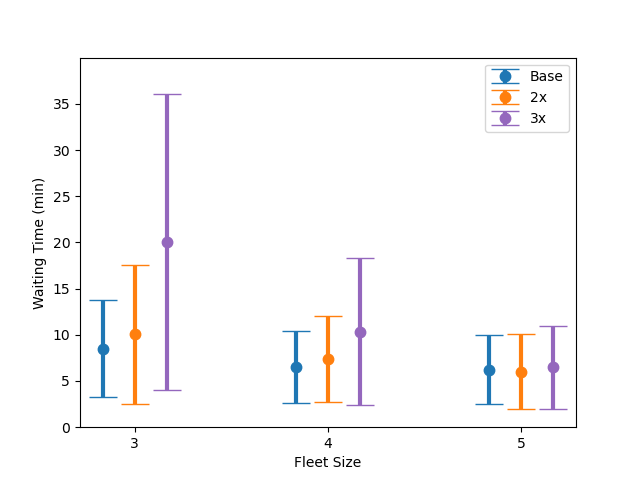}
        \caption{Belvedere - SFL $III$}
        \label{fig:wait-westatlanta-rider-av}
    \end{subfigure}
    \begin{subfigure}[]{0.45\textwidth}
        \includegraphics[width=\textwidth]{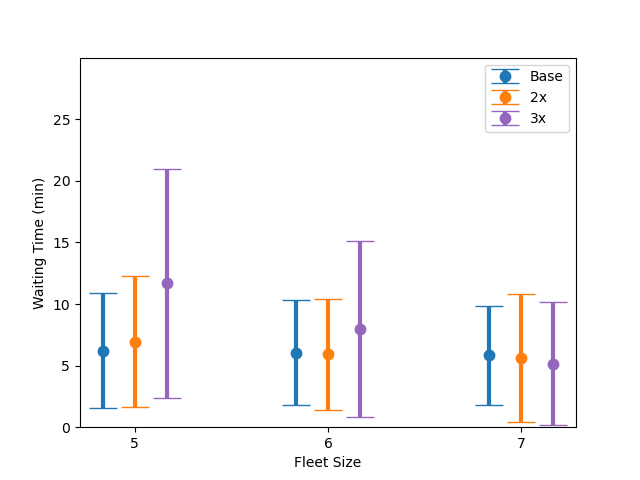}
        \caption{West Atlanta - SFL $III$}
        \label{fig:wait-belvedere-rider-av}
    \end{subfigure}
\caption{Average Waiting Time Per Trip Across Ridership Levels}
\label{fig:avg-wait-time-per-trip-rider}
\end{figure}

\subsection{Analysis of Waiting and Travel Times}
\label{subsec:wait-time-trips}
Figures~\ref{fig:wait-belvedere-mode-base} and~\ref{fig:wait-westatlanta-mode-base} present the average waiting time per \revision{trip} across different SFLs and fleet sizes at the base ridership level for Belvedere and West Atlanta, respectively. Dots indicate mean values and bars represent one standard deviation from the mean. Evidently, under the same ridership scenario, increasing the fleet size consistently reduces \revision{average} waiting times in both zones, as more shuttles can accommodate demand simultaneously. \revision{Even though all trips are guaranteed service, the vast majority of trips experience reasonable waiting times. Across all SFLs and fleet sizes, the 95th percentile waiting times are 19.35 minutes in Belvedere and 14.89 minutes in West Atlanta, demonstrating that users can be served conveniently under the base ridership scenario.}

In the context of SFLs, SFLs $I$ and $II$, which involve human drivers, the average wait times range between 7-10 minutes for Belvedere and 6-8 minutes for West Atlanta. As expected, eliminating driver response delays in SFL $II$ yields slightly lower waiting times than SFL $I$. While the average improvement is \revision{marginal}, eliminating driver responses prevents instances of excessively long waiting times that can occur with human drivers, as previously illustrated in Figure~\ref{fig:response_time_distribution}. \revision{In both zones, the introduction of SAVs further reduces average passenger waiting times.} \revision{Specifically, average waiting times decrease from SFL $II$ to $III$, representing the transition from human-driven shuttles to SAVs.}
\revision{Although the reduction in mean waiting times is limited, as indicated by overlapping standard deviation ranges, the overall decrease in variability suggests that more riders consistently experience shorter waiting times.} The benefits are further \revision{evidenced} in SFL $IV$, which relocates idle stops toward frequent origin points. This indicates that combining SAVs with demand-responsive idle stops offers \revision{greater improvements} in service quality and operational efficiency.

While SAVs may introduce additional detours to accommodate on-demand requests and reduce passenger waiting times, these detours do not compromise overall travel times for trips. Figures~\ref{fig:travel-belvedere-mode-base} and~\ref{fig:travel-westatlanta-mode-base} show the average travel time per \revision{trip} across different SFLs and fleet sizes at the base ridership level for the Belvedere and West Atlanta zones, respectively. The results indicate that travel times remain largely unchanged \revision{across all SFLs}, suggesting that any additional detours from more dynamic pickups introduced by SAV operations do not negatively impact trip durations. This outcome demonstrates that the RTDARS-SAV dispatching model effectively routes shuttles to balance detour flexibility with service efficiency, thereby maintaining high service quality.

Figure~\ref{fig:avg-wait-time-per-trip-rider} illustrates the effects of higher ridership on the average wait times for SFLs $I$ and $III$, respectively, in both Belvedere and West Atlanta. Across both SFLs, the trends are consistent: at lower fleet sizes, 3x ridership leads to pronounced increases in waiting times, reflecting an over-saturated system. Large standard deviations suggest that some riders are severely impacted. However, expanding the fleet size alleviates these effects by improving the system’s capacity to serve additional passengers. Across all SFLs and fleet sizes at 3x ridership, the 95th percentile waiting times reach 53.05 minutes in Belvedere and 37.13 minutes in West Atlanta, which are considerably high. However, given the elevated demand levels, these longer waiting times could be substantially reduced by increasing the fleet size, as additional vehicles would improve service capacity under higher ridership conditions.

\subsection{Analysis of Shuttle Distances}
\label{subsec:empty-distance}

\begin{figure}[!ht]
    \centering
    \begin{subfigure}[]{0.45\textwidth}
        \includegraphics[width=\textwidth]{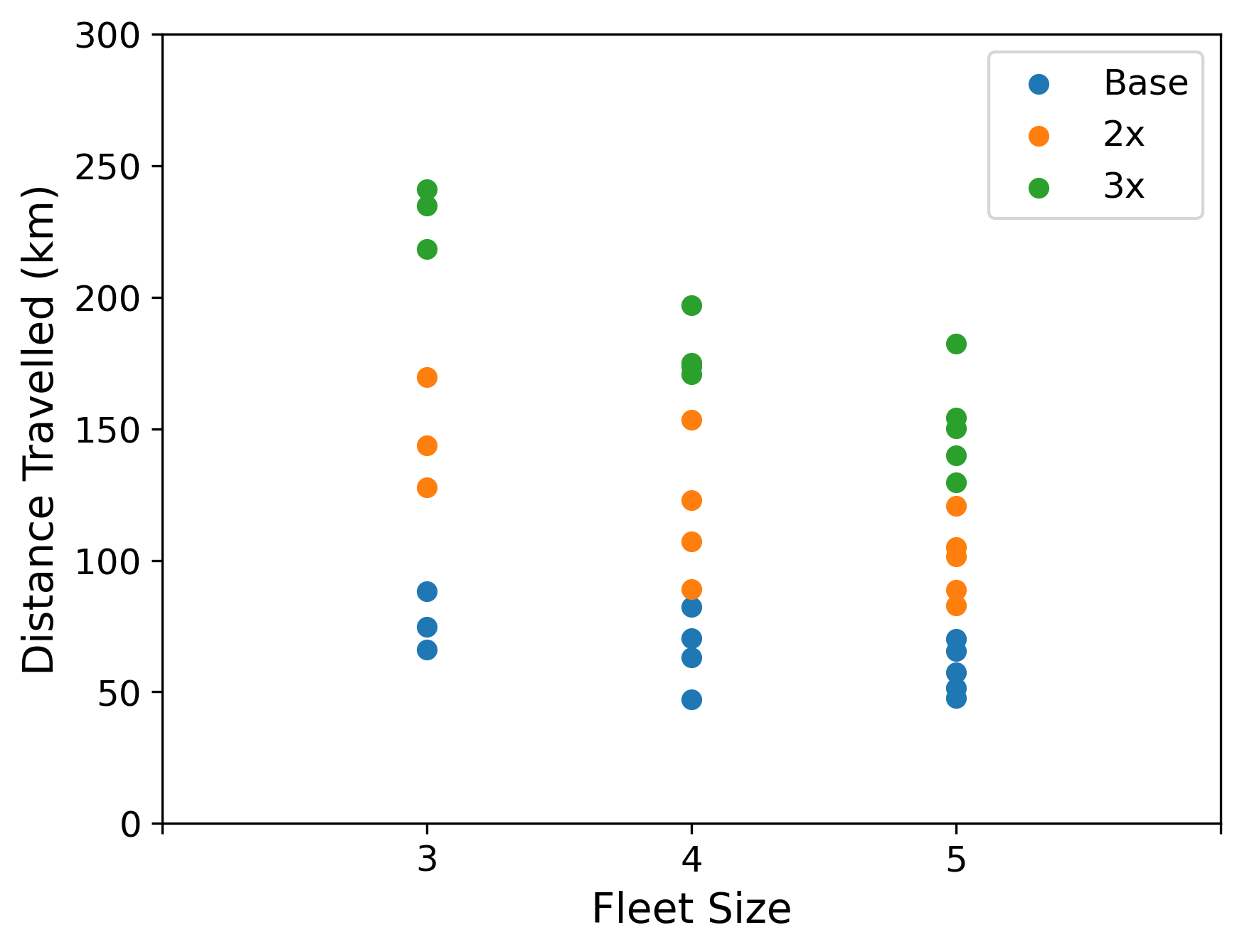}
        \caption{SFL $I$}
        \label{fig:shuttle-distances-belvedere-db}
    \end{subfigure}
    \begin{subfigure}[]{0.45\textwidth}
        \includegraphics[width=\textwidth]{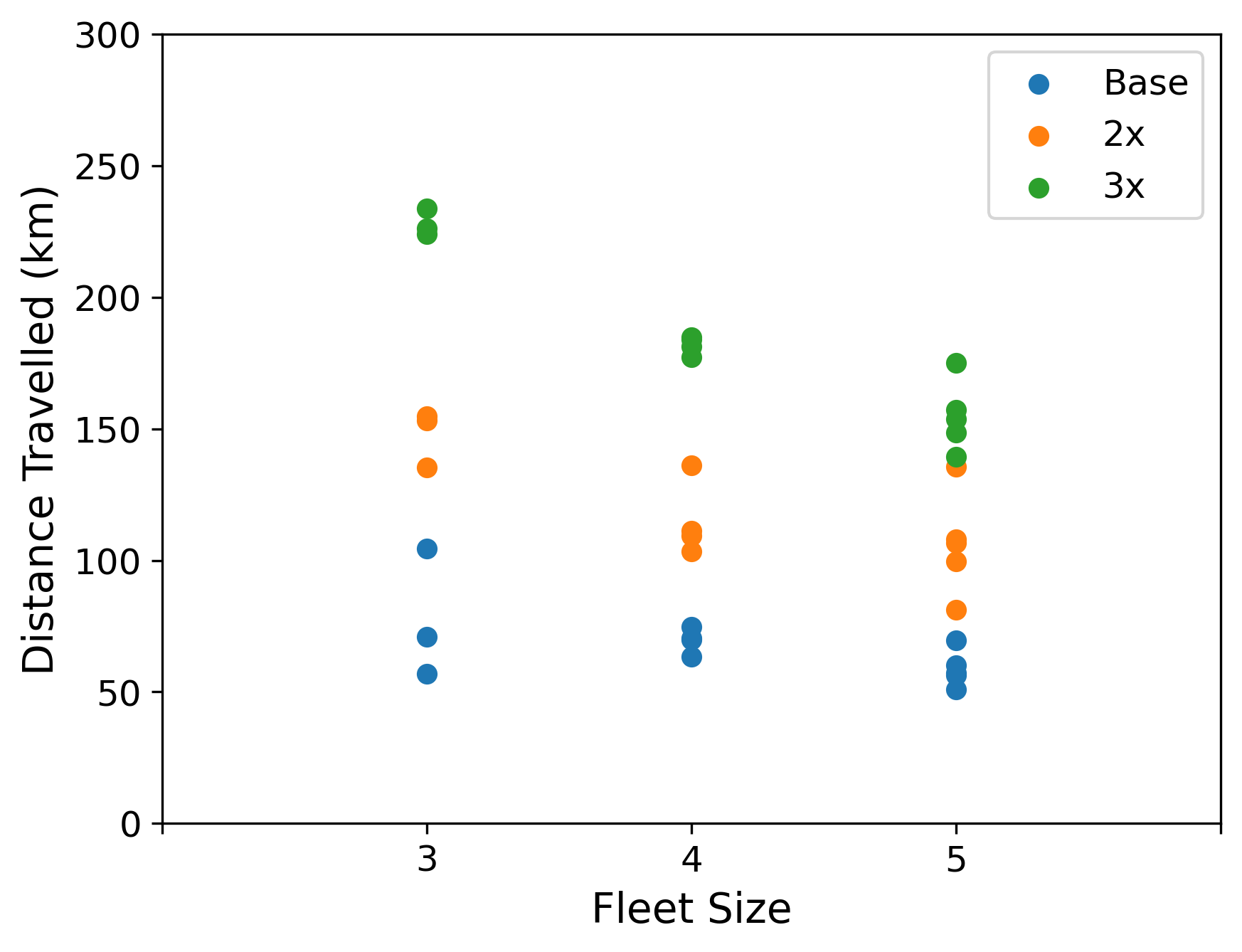}
        \caption{SFL $II$}
        \label{fig:shuttle-distances-belvedere-}
    \end{subfigure}
    \begin{subfigure}[]{0.45\textwidth}
        \includegraphics[width=\textwidth]{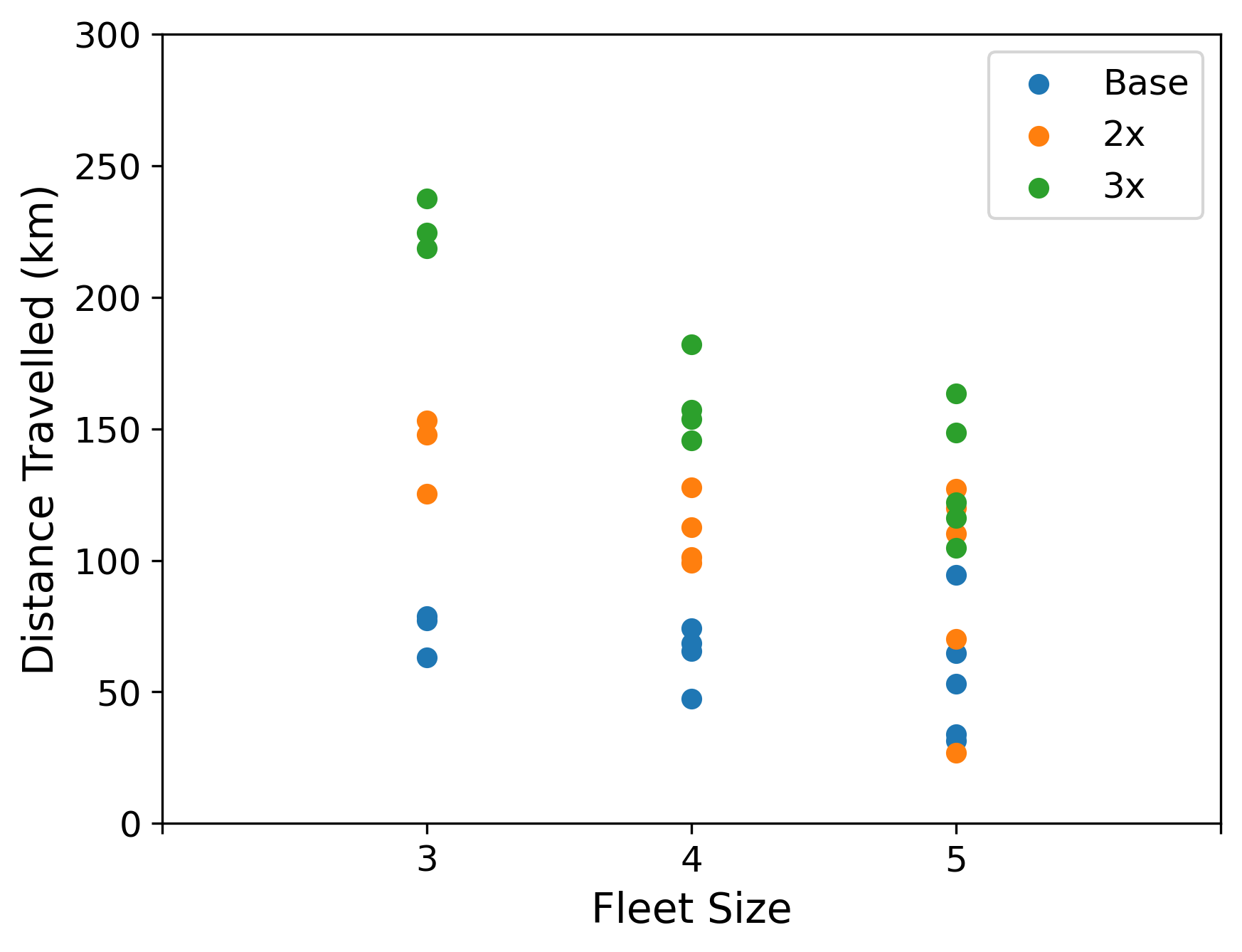}
        \caption{SFL $III$}
        \label{fig:shuttle-distances-belvedere-av}
    \end{subfigure}
    \begin{subfigure}[]{0.45\textwidth}
        \includegraphics[width=\textwidth]{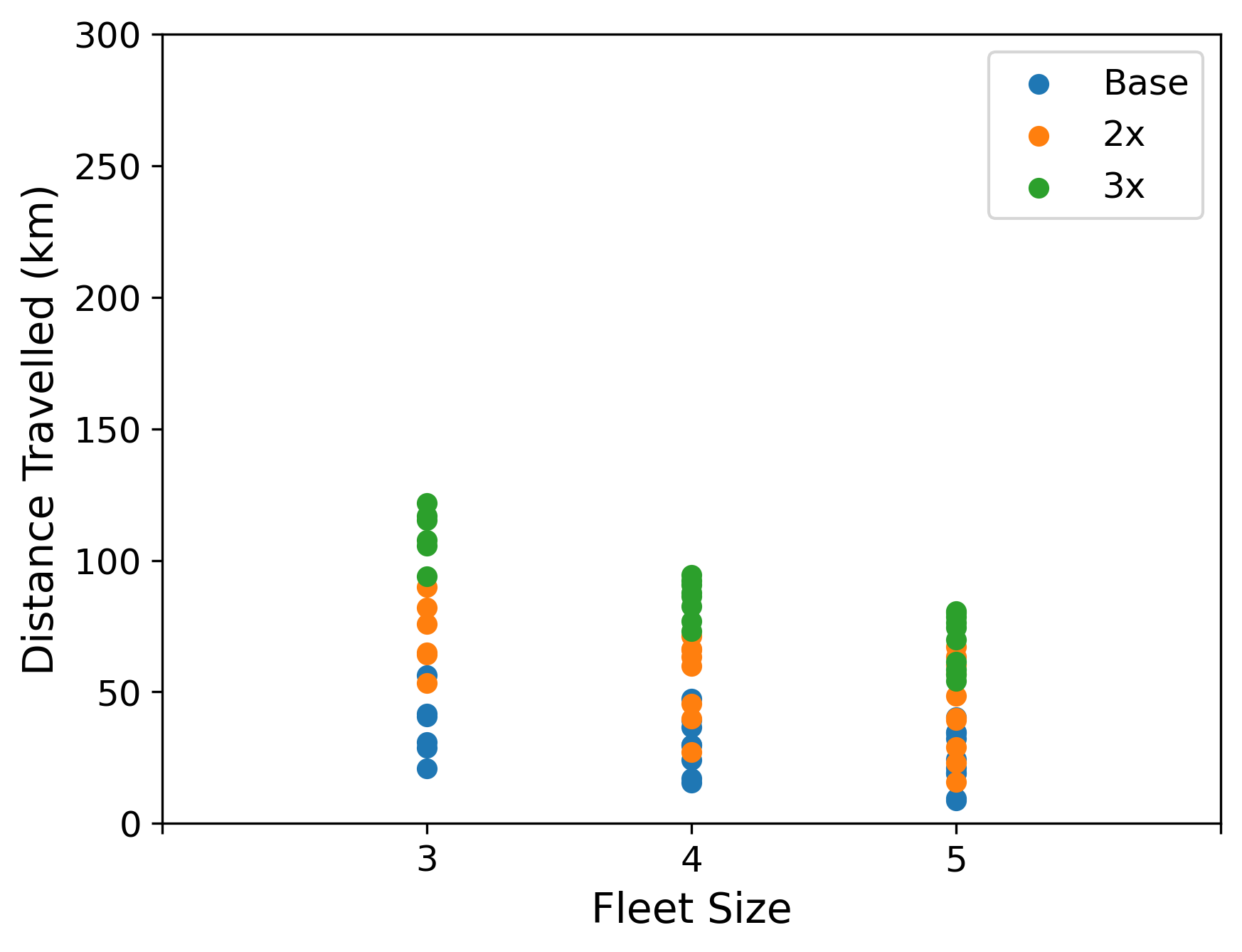}
        \caption{SFL $IV$}
        \label{fig:shuttle-distances-belvedere-avir}
    \end{subfigure}
\caption{Distance Traveled by Each Shuttle in Belvedere}
\label{fig:shuttle-distances-belvedere}
\end{figure}

\begin{figure}[!ht]
    \centering
    \begin{subfigure}[]{0.45\textwidth}
        \includegraphics[width=\textwidth]{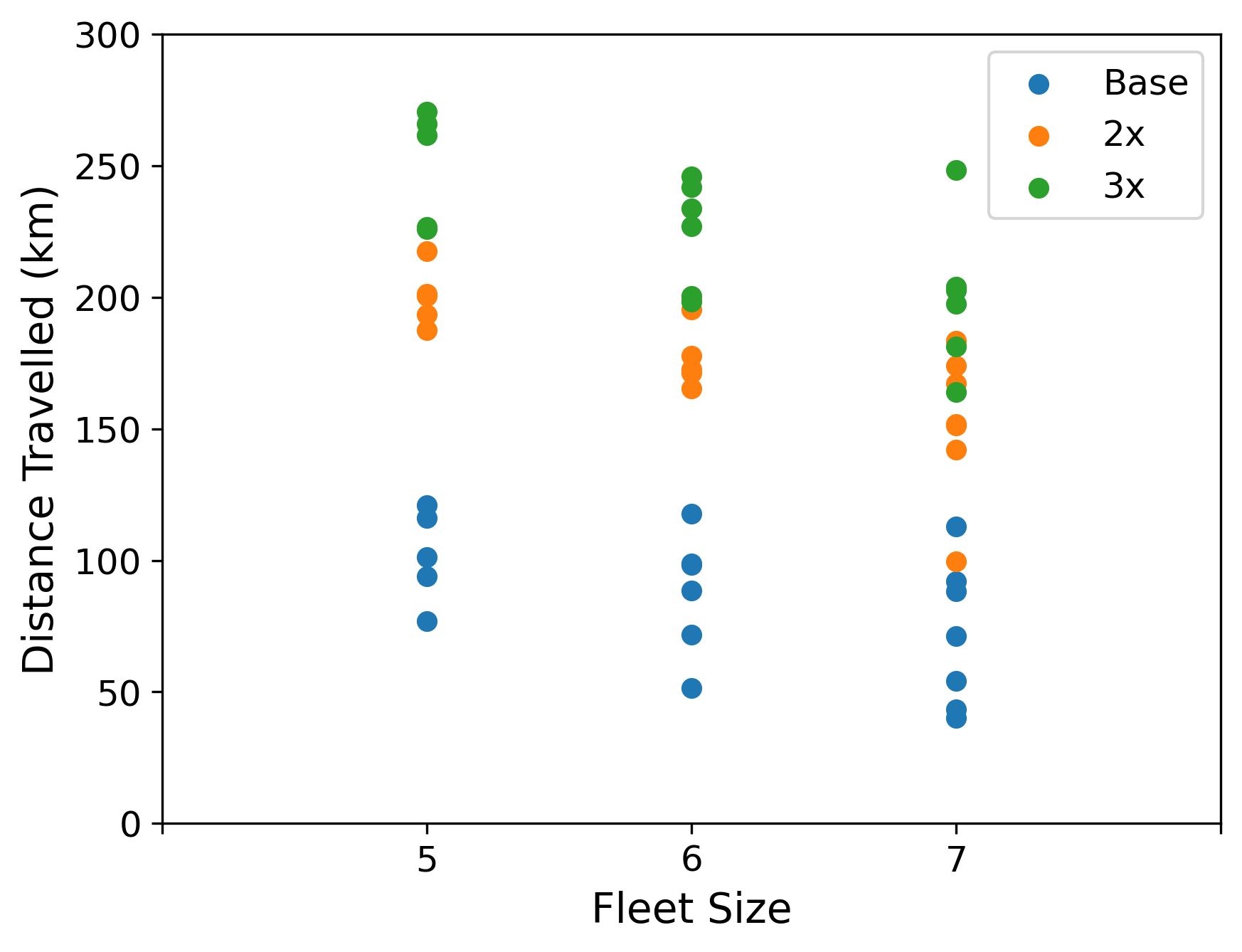}
        \caption{SFL $I$}
        \label{fig:shuttle-distances-westatlanta-db}
    \end{subfigure}
    \begin{subfigure}[]{0.45\textwidth}
        \includegraphics[width=\textwidth]{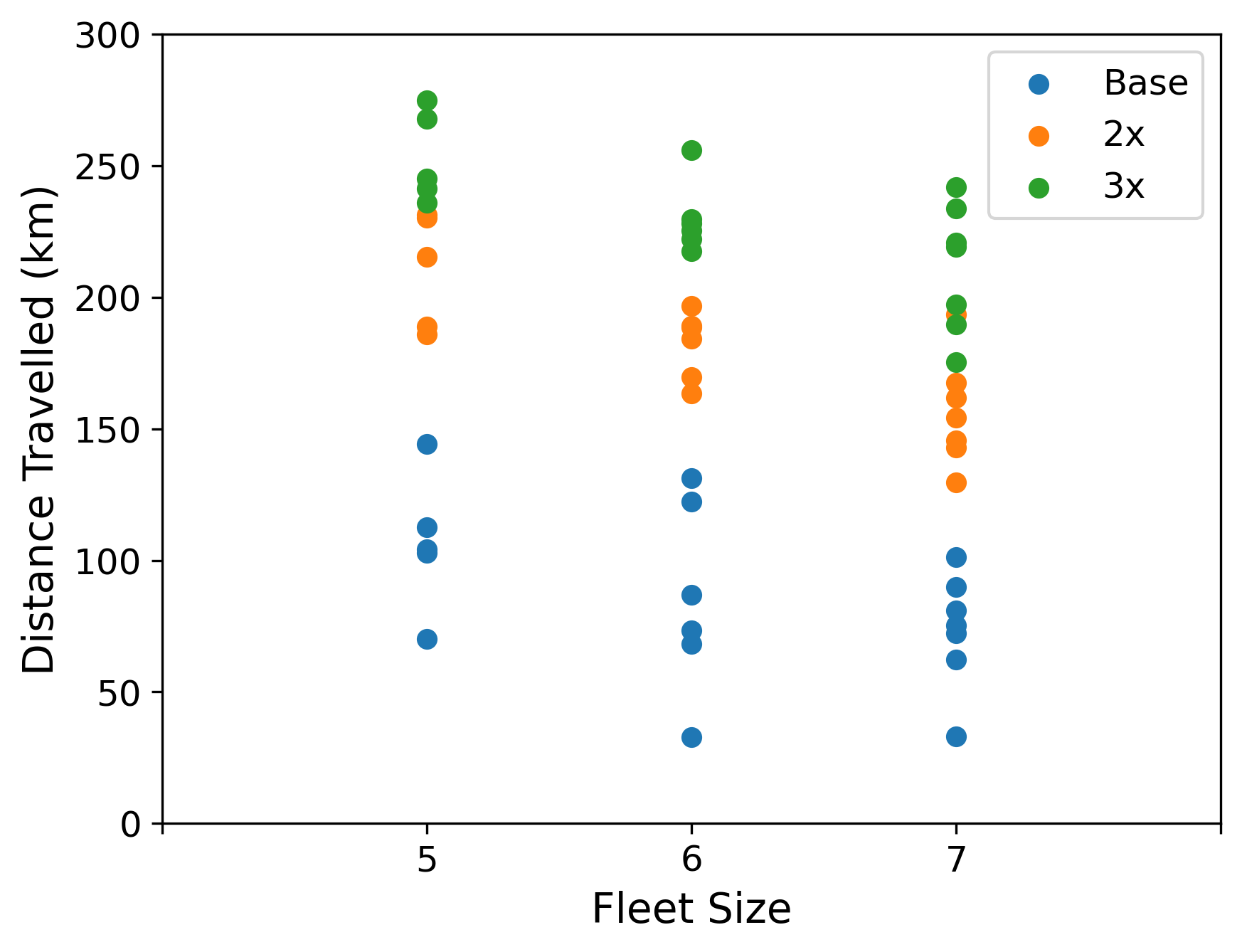}
        \caption{SFL $II$}
        \label{fig:shuttle-distances-westatlanta-}
    \end{subfigure}
    \begin{subfigure}[]{0.45\textwidth}
        \includegraphics[width=\textwidth]{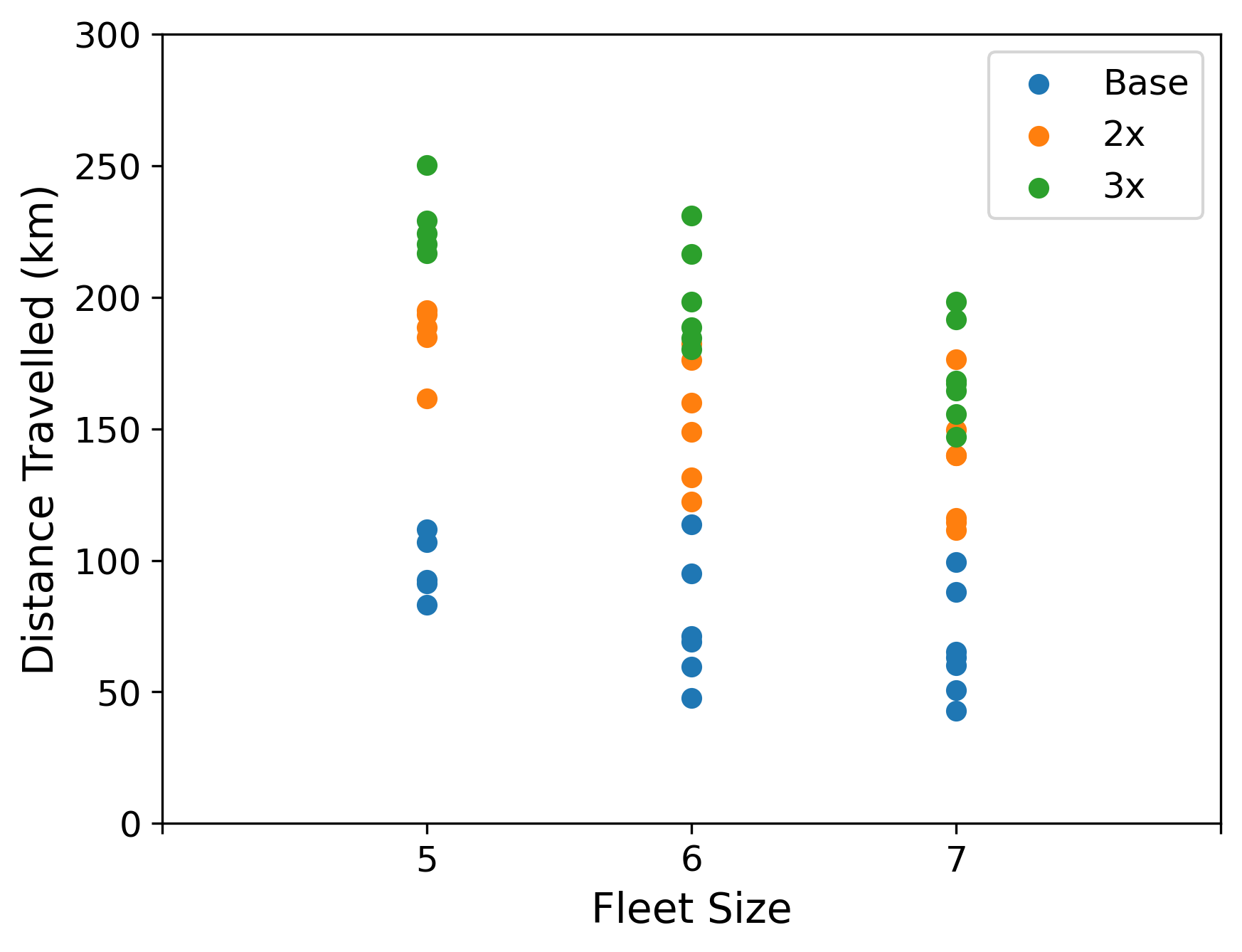}
        \caption{SFL $III$}
        \label{fig:shuttle-distances-westatlanta-av}
    \end{subfigure}
    \begin{subfigure}[]{0.45\textwidth}
        \includegraphics[width=\textwidth]{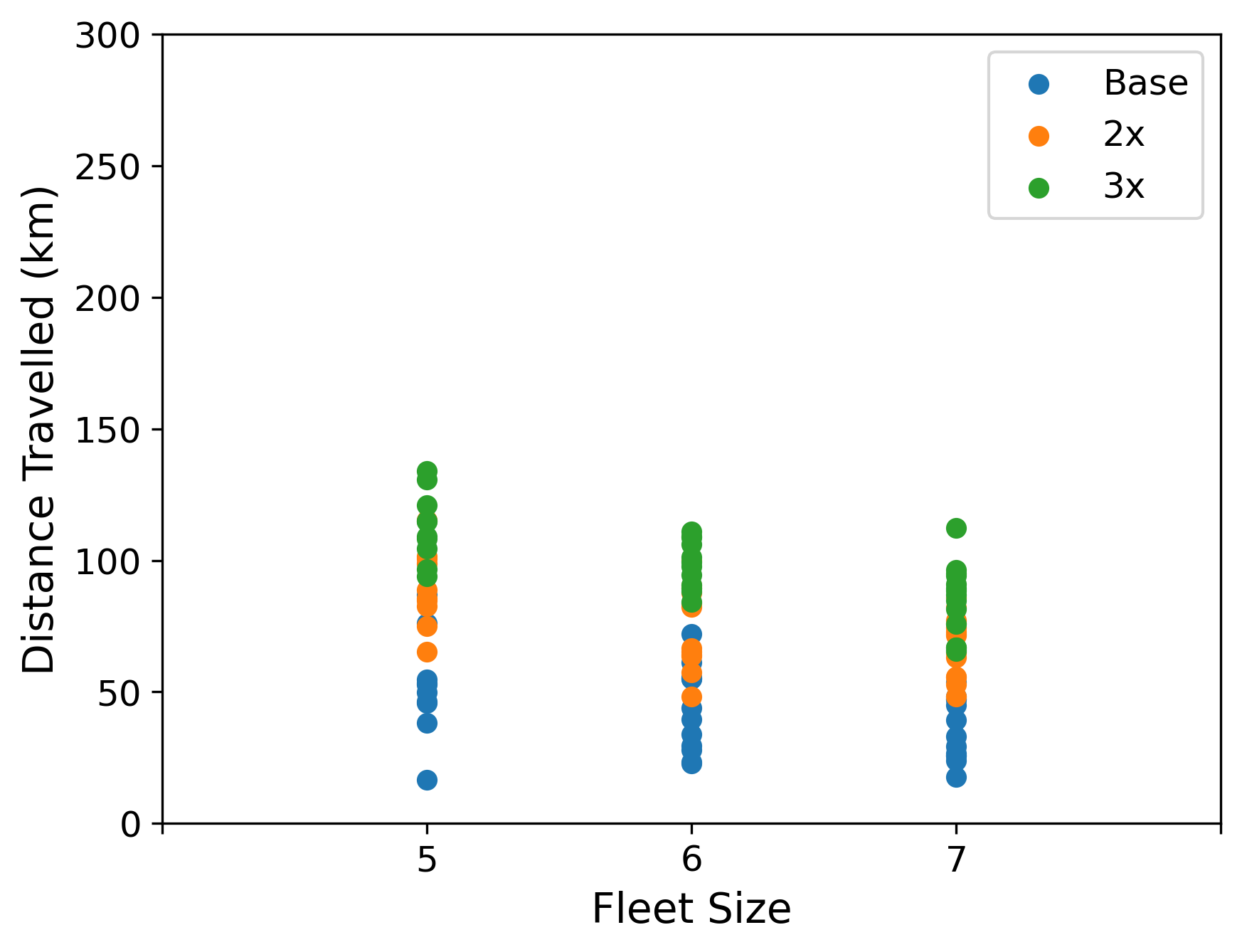}
        \caption{SFL $IV$}
        \label{fig:shuttle-distances-westatlanta-avir}
    \end{subfigure}
\caption{Distance Traveled by Each Shuttle in West Atlanta}
\label{fig:shuttle-distances-westatlanta}
\end{figure}

Figures~\ref{fig:shuttle-distances-belvedere} and~\ref{fig:shuttle-distances-westatlanta} present the total kilometers traveled by each shuttle in Belvedere and West Atlanta, respectively, across different SFLs. Each point represents an individual shuttle under a particular combination of fleet size, ridership, and SFL. As expected in both zones, increasing the fleet size generally reduces the distance traveled by shuttles, as a larger number of vehicles share the same overall ridership. Conversely, the total distance traveled increases as ridership levels increase, reflecting the need to cover greater distances to accommodate more trip requests.

The sustainability benefits of SAVs are evident from the travel distance insights. Shuttles of SFL $III$ travel fewer kilometers on average compared to shuttles of SFLs $I$ and $II$, primarily due to their ability to dynamically re-route and serve requests en route. For SFLs $I$ and $II$, shuttles with human drivers may need to backtrack routes to accommodate requests they could not serve while driving, resulting in greater kilometers traveled. Notably, SFL $IV$ scenarios yield the shortest distance traveled. Since SFL $IV$ incorporates strategic idling at locations with higher anticipated demand, shuttles are positioned closer to future requests, minimizing travel distances. This underscores the sustainability advantages of SFL $IV$ and the value of strategically planning idle location placement in response to spatial and temporal demand patterns.

The shuttle distances also validate the effectiveness of both RTDARS and RTDARS-SAV dispatching models in effectively dispatching and managing shuttle fleets across all scenarios. Under any scenario, all shuttles exhibit relatively similar travel distances, indicating balanced utilization of shuttle fleets. 

It is noteworthy that across all scenarios, shuttles traveled no more than approximately 300 kilometers throughout the entire simulation day. Assuming that SAVs are electric and begin the day with a full charge, this distance remains well within the typical range of modern electric minivans, which can exceed 350 kilometers on a single charge \citep{Mercedes,2025Volkswagen}.

Tables~\ref{tab:empty-miles-belvedere} and~\ref{tab:empty-miles-westatlanta} show the empty kilometers for the shuttle fleets. The analysis indicates that SAVs have substantial potential to reduce not only travel kilometers traveled, but also total empty kilometers traveled by the fleet, especially as the fleet size increases. In Belvedere, SFLs $III$ and $IV$ consistently result in lower total kilometers traveled across all fleet sizes and ridership levels compared to SFLs $I$ and $II$. Moreover, the kilometer savings generally grow at an increasing rate with both fleet size and ridership level. A similar pattern is observed in West Atlanta, where SFLs $III$ and $IV$ demonstrate consistent reductions in empty kilometers over SFLs $I$ and $II$.

\begin{table}[!ht]
\centering
\begin{tabular}{lrrr lrrr lrrr}
\toprule
\multirow{2}{*}{SFL} & \multicolumn{3}{c}{$|N| = 39$: 1x (Base)}       &  & \multicolumn{3}{c}{$|N| = 78$: 2x}         &  & \multicolumn{3}{c}{$|N| = 117$: 3x}         \\ \cmidrule{2-4} \cmidrule{6-8} \cmidrule{10-12}
\multicolumn{1}{c}{} & $|V| = 3$      & $|V| = 4$     & $|V| = 5$      &  & $|V| = 3$      & $|V| = 4$     & $|V| = 5$    &  & $|V| = 3$      & $|V| = 4$     & $|V| = 5$     \\ \midrule
$I$                           & 98.4   & 132.2  & 161.8  &  & 200.1  & 226.8 & 257.4 &  & 304.5  & 334.2 & 366.1 \\
$II$                           & 104.1  & 147.6  & 163.5  &  & 207.4  & 222.0 & 292.3 &  & 310.5  & 339.2 & 387.6 \\
$III$                           & 88.6   & 124.9  & 147.1  &  & 182.5  & 201.9 & 213.4 &  & 299.1  & 252.5 & 264.5 \\
$IV$                         & 88.6   & 105.5  & 109.6  &  & 186.5  & 192.6 & 194.7 &  & 285.8  & 296.5 & 304.8 \\ 
\bottomrule
\end{tabular}
\caption{Total Empty Kilometers for Fleets in Belvedere, $|N|$ is the number of trip requests and $|V|$ is the fleet size}
\label{tab:empty-miles-belvedere}
\end{table}

\begin{table}[!ht]
\centering
\begin{tabular}{lrrr lrrr lrrr}
\toprule
\multirow{2}{*}{SFL} & \multicolumn{3}{c}{$|N| = 82$: 1x (Base)}       &  & \multicolumn{3}{c}{$|N| = 164$: 2x}         &  & \multicolumn{3}{c}{$|N| = 246$: 3x}         \\ \cmidrule{2-4} \cmidrule{6-8} \cmidrule{10-12}
\multicolumn{1}{c}{} & $|V| = 5$      & $|V| = 6$     & $|V| = 7$      &  & $|V| = 5$      & $|V| = 6$     & $|V| = 7$    &  & $|V| = 5$      & $|V| = 6$     & $|V| = 7$     \\ \midrule
$I$                           & 276.2  & 285.3  & 264.8  &  & 490.5  & 569.7 & 560.6 &  & 552.1  & 629.0 & 685.8 \\
$II$                           & 293.2  & 273.8  & 273.9  &  & 548.8  & 589.6 & 587.6 &  & 564.8  & 660.2 & 752.8 \\
$III$                           & 236.9  & 211.1  & 224.6  &  & 420.3  & 415.6 & 440.6 &  & 446.9  & 486.0 & 478.2 \\
$IV$                         & 272.0  & 271.7  & 281.0  &  & 405.4  & 393.5 & 442.5 &  & 435.7  & 466.3 & 471.8 \\ \hline
\end{tabular}
\caption{Total Empty Kilometers for Fleets in West Atlanta}
\label{tab:empty-miles-westatlanta}
\end{table}

Even as ridership increases, fleets operating under SFLs $III$ and $IV$ maintain shorter empty kilometers than SFLs $I$ and $II$. While total empty kilometer increases for all SFLs with increasing ridership, the consistency of SAVs in reducing kilometers driven signifies a notable operational benefit. In particular, at higher ridership levels and larger fleet sizes, SFLs $III$ and $IV$ show increasingly significant reductions in empty kilometers compared to SFLs $I$ and $II$. These consistent patterns across both the Belvedere and West Atlanta zones highlight the scalability and robustness of the SAV advantage in minimizing unproductive shuttle movement.

\subsection{Analysis of Ridesharing Rates}
\label{subsec:ridesharing}

\begin{figure}[!ht]
    \centering
    \begin{subfigure}[]{0.45\textwidth}
        \includegraphics[width=\textwidth]{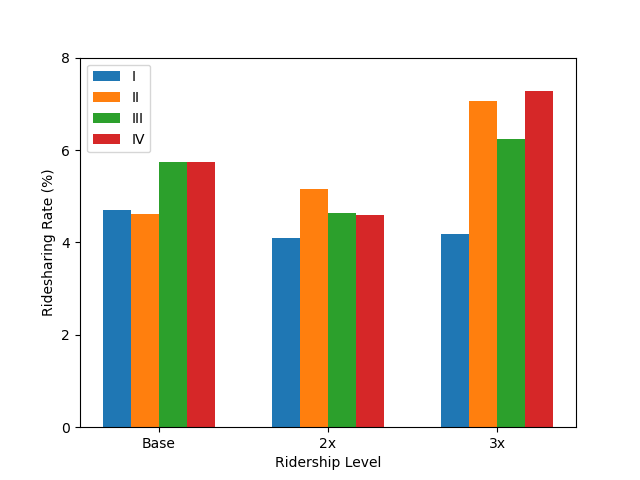}
        \caption{Belvedere - Fleet Size: 3}
        \label{fig:avg-ridesharing-distance-belvedere-fleet-3}
    \end{subfigure}
    \begin{subfigure}[]{0.45\textwidth}
        \includegraphics[width=\textwidth]{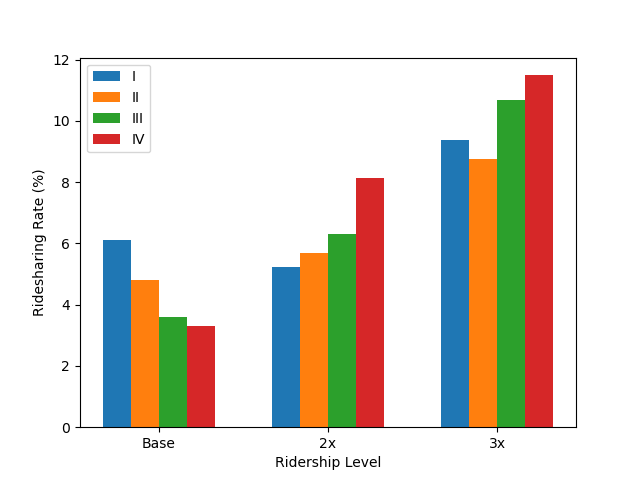}
        \caption{West Atlanta - Fleet Size: 5}
        \label{fig:avg-ridesharing-distance-westatlanta-fleet-5}
    \end{subfigure}
    \begin{subfigure}[]{0.45\textwidth}
        \includegraphics[width=\textwidth]{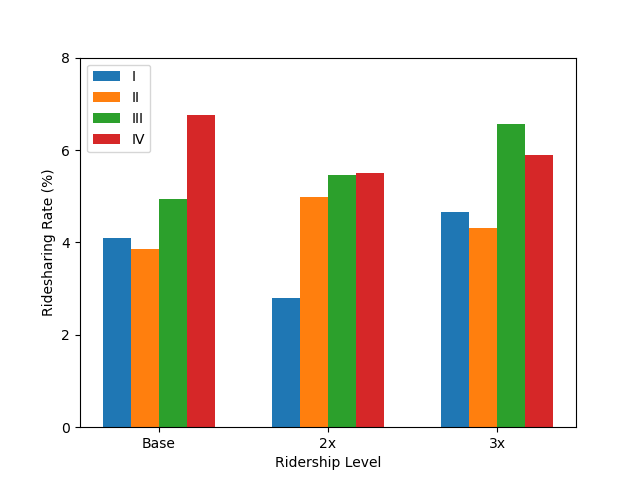}
        \caption{Belvedere - Fleet Size: 4}
        \label{fig:avg-ridesharing-distance-belvedere-fleet-4}
    \end{subfigure}
    \begin{subfigure}[]{0.45\textwidth}
        \includegraphics[width=\textwidth]{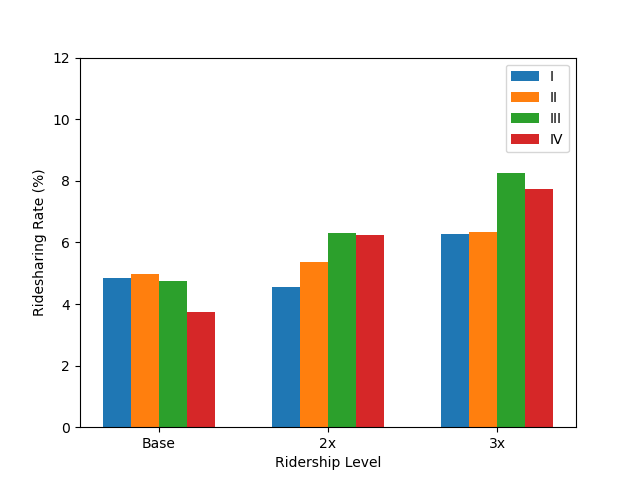}
        \caption{West Atlanta - Fleet Size: 6}
        \label{fig:avg-ridesharing-distance-westatlanta-fleet-6}
    \end{subfigure}
    \begin{subfigure}[]{0.45\textwidth}
        \includegraphics[width=\textwidth]{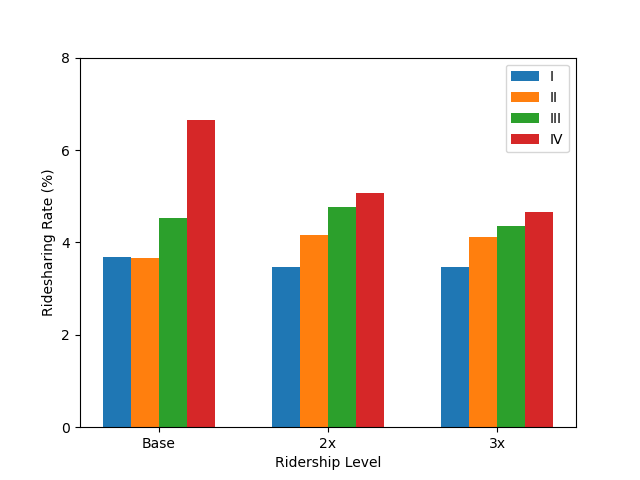}
        \caption{Belvedere - Fleet Size: 5}
        \label{fig:avg-ridesharing-distance-belvedere-fleet-5}
    \end{subfigure}
    \begin{subfigure}[]{0.45\textwidth}
        \includegraphics[width=\textwidth]{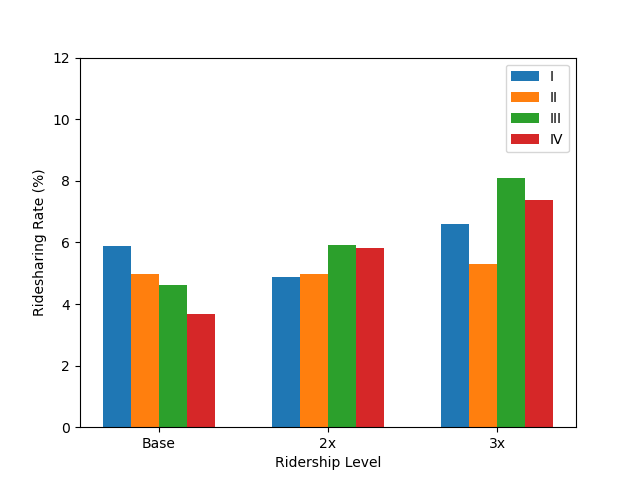}
        \caption{West Atlanta - Fleet Size: 7}
        \label{fig:avg-ridesharing-distance-westatlanta-fleet-7}
    \end{subfigure}
\caption{Average Ridesharing Rates}
\label{fig:avg-ridesharing-rates}
\end{figure}

Figure~\ref{fig:avg-ridesharing-rates} presents the ridesharing rates for Belvedere and West Atlanta for different fleet sizes. The ridesharing rate is defined as the proportion of total kilometers traveled by the fleet while carrying more than one passenger. As expected, ridesharing rates decrease with larger fleet sizes as a greater number of shuttles allows for more flexible routing and reduces the need to pool trips together for the same trips. At lower fleet sizes, ridesharing rates generally increase with higher demand, as a larger pool of trips raises the likelihood of shared trips. At higher fleet sizes, however, ridesharing rates tend to stabilize, as the available fleet can accommodate increased demand without requiring additional pooling.

Between Belvedere and West Atlanta, ridesharing patterns differ, and overall, the impact of SFLs on ridesharing is not evident. In Belvedere, ridesharing rates are consistently higher for SAVs across all ridership and fleet size scenarios. However, in West Atlanta, SAVs exhibit lower ridesharing rates at lower ridership levels but higher rates compared to human-driven shuttles as demand increases. This indicates that ridesharing rates are highly dependent on spatial and temporal factors, which in turn shape how different SFLs affect the extent of shared travel.

Overall, ridesharing levels were relatively low. However, this outcome is expected given the geo-fenced nature of the zones, which cover small areas and consequently result in shorter trips that must be completed within a zone. Furthermore, a separate ODMTS study similarly found that ridesharing among human drivers exhibited moderate ridesharing rates, which is consistent with the findings of this study \citep{auad2021resiliency}. These insights collectively suggest that a modest fleet size has the potential to serve even more users effectively.

\subsection{Analysis of Operational Costs}

Tables~\ref{tab:cost-per-rider-ridership-belvedere} and~\ref{tab:cost-per-rider-ridership-westatlanta} present the sensitivity analysis of cost per \revision{user} across varying operational costs per shuttle per hour for Belvedere and West Atlanta, respectively. Cost per \revision{user} was used instead of cost per \revision{trip} because \revision{MARTA fares are paid by individual users}. The cost per shuttle per hour parameters are hypothetical and are used solely for sensitivity analysis. The cost per shuttle per hour was multiplied by the total hours of operation and the number of shuttles, and then divided by the number of \revision{users} to obtain the cost per \revision{user}. Regardless of the SFL, the cost per \revision{user} remains consistent because the number of riders served does not change across SFLs. Moreover, labor costs are assumed to remain consistent across SFLs. While SAVs eliminate the need for human drivers, operational labor may still be required to monitor SAVs while in service\revision{, particularly in the early stages of deployment}. Recent developments in SAV-based mobility services have shown that operators often include attendants on board to greet passengers, provide service information, and intervene manually if necessary. In addition, SAV operators increasingly rely on command center agents who remotely monitor vehicles and provide assistance when issues arise outside the operational domain \citep{Beep}.

Evidently, the cost per \revision{user} increases with larger fleet sizes, as the higher operational costs outpace the corresponding increase in ridership. However, this higher cost per \revision{user} is accompanied by an improvement in service quality from the expanded fleet capacity. The results also indicate that the cost per \revision{user} decreases substantially as ridership increases. 

For MARTA Reach shuttles with a capacity of eight passengers, the average cost per shuttle per hour is estimated between \$50 to \$70 \citep{vanhentenryck2023reach}. In contrast, ridesharing services such as \revision{Uber} typically use compact vehicles such as sedans, which have a capacity of four passengers and a per-vehicle operating cost of approximately \$26.35 per vehicle hour, including labor and operating expenses \citep{helling_stephen_hempler_paul_williams_savage_tom_2023}. Therefore, minivan shuttles with a capacity of six passengers are estimated to have an average operating cost between \$35 and \$50 per hour. This results in a cost per \revision{user} between \$31.74 to \$75.58 in Belvedere and \$25.56 and \$51.12 in West Atlanta at base ridership levels. As public transit is highly subsidized, fare revenues cover only a portion of operating expenses. MARTA fares specifically accounted for only 12.2\% of the total operating expenses in 2022 \citep{FTA2022-NationalTransitDatabase}, with a fixed fare of \$2.50 per \revision{user}. To maintain an equivalent fare-to-cost ratio, the subsequent fares would range from \$3.87 to \$9.22 per \revision{user} in Belvedere and \$3.11 to \$6.24 in West Atlanta. At
higher demand levels, the cost per \revision{user} lies between \$11.19 and \$26.64 in Belvedere and \$8.43 and \$16.85 in West Atlanta, with corresponding adjusted fares ranging from \$1.37 to \$3.25 in Belvedere and \$1.03 \revision{to} \$2.06 in West Atlanta, which are comparable to, and potentially less expensive than, MARTA's current transit fare. This suggests that integrating on-demand minivan shuttles to complement bus service could yield substantial cost savings for public transit agencies, with increasing returns at scale with demand. This is particularly relevant, as the pilot data showed a continuous increase in demand toward its conclusion, indicating the potential for more cost-efficient operations.

Additional scenarios with higher cost per shuttle per hour values are provided in Appendix~\ref{sec:additional-operational-costs}.

\begin{table}[!ht]
\centering
\begin{tabular}{lrrr lrrr lrrr}
\toprule
\multicolumn{1}{c}{\multirow{2}{*}{\begin{tabular}[c]{@{}c@{}} Shuttle \\ Cost \end{tabular}}} & \multicolumn{3}{c}{$|N| = 39$: 1x (Base)}       &  & \multicolumn{3}{c}{$|N| = 78$: 2x}         &  & \multicolumn{3}{c}{$|N| = 117$: 3x}         \\ \cmidrule{2-4} \cmidrule{6-8} \cmidrule{10-12}
\multicolumn{1}{c}{} & $|V| = 3$      & $|V| = 4$     & $|V| = 5$      &  & $|V| = 3$      & $|V| = 4$     & $|V| = 5$    &  & $|V| = 3$      & $|V| = 4$     & $|V| = 5$     \\ \midrule

20                                                                                                             & 18.14     & 24.19     & 30.23     &  & 9.51       & 12.68     & 15.85   &  & 6.39     & 8.52    & 10.66
  \\
25                                                                                                             & 22.67     & 30.23     & 37.79     &  & 11.89       & 15.85     & 19.82   &  & 7.99     & 10.66    & 13.32
  \\
30                                                                                                             & 27.21     & 36.28     & 45.35     &  & 14.27       & 19.02     & 23.78   &  & 9.59     & 12.79    & 15.98
  \\
35                                                                                                             & 31.74     & 42.33     & 52.91     &  & 16.65       & 22.20     & 27.74   &  & 11.19     & 14.92    & 18.65
  \\
40                                                                                                             & 36.28     & 48.37     & 60.47     &  & 19.02       & 25.37     & 31.71   &  & 12.79     & 17.05    & 21.31
  \\
45                                                                                                             & 40.81     & 54.42     & 68.02     &  & 21.40       & 28.54     & 35.67   &  & 14.39     & 19.18    & 23.98
  \\
50                                                                                                             & 45.35     & 60.47     & 75.58     &  & 23.78       & 31.71     & 39.63   &  & 15.98     & 21.31    & 26.64   \\ \hline
\end{tabular}
\caption{Cost (\$) per \revision{user} in Belvedere Across Ridership Levels. Shuttle costs are in \$/hour.}
\label{tab:cost-per-rider-ridership-belvedere}
\end{table}

\begin{table}[!ht]
\centering
\begin{tabular}{lrrr lrrr lrrr}
\toprule
\multicolumn{1}{c}{\multirow{2}{*}{\begin{tabular}[c]{@{}c@{}} Shuttle \\ Cost \end{tabular}}} & \multicolumn{3}{c}{$|N| = 82$: 1x (Base)}       &  & \multicolumn{3}{c}{$|N| = 164$: 2x}         &  & \multicolumn{3}{c}{$|N| = 246$: 3x}         \\ \cmidrule{2-4} \cmidrule{6-8} \cmidrule{10-12}
\multicolumn{1}{c}{} & $|V| = 5$      & $|V| = 6$     & $|V| = 7$      &  & $|V| = 5$      & $|V| = 6$     & $|V| = 7$    &  & $|V| = 5$      & $|V| = 6$     & $|V| = 7$     \\ \midrule
20                                                                                                             & 14.61    & 17.53    & 20.45    &  & 7.14     & 8.57    & 10.00   &  & 4.81     & 5.78    & 6.74   \\
25                                                                                                             & 18.26    & 21.91    & 25.56    &  & 8.93     & 10.71    & 12.50   &  & 6.02     & 7.22    & 8.43   \\
30                                                                                                             & 21.91    & 26.29    & 30.67    &  & 10.71     & 12.86    & 15.00   &  & 7.22     & 8.67    & 10.11   \\
35                                                                                                             & 25.56    & 30.67    & 35.79    &  & 12.50     & 15.00    & 17.50   &  & 8.43     & 10.11    & 11.80   \\
40                                                                                                             & 29.21    & 35.06    & 40.90    &  & 14.29     & 17.14    & 20.00   &  & 9.63     & 11.56    & 13.48   \\
45                                                                                                             & 32.87    & 39.44    & 46.01    &  & 16.07     & 19.29    & 22.50   &  & 10.83     & 13.00    & 15.17   \\
50                                                                                                             & 36.52    & 43.82    & 51.12    &  & 17.86     & 21.43    & 25.00   &  & 12.04     & 14.44    & 16.85   \\ \hline
\end{tabular}
\caption{Cost (\$) per \revision{user} in West Atlanta Across Ridership Levels. Shuttle costs are in \$/hour.}
\label{tab:cost-per-rider-ridership-westatlanta}
\end{table}
\section{Conclusion}
\label{sec:conclusion}

This paper investigated the impact of SAVs in microtransit systems, alongside other operational features. To implement this study, a RTDARS modeling framework was employed and enhanced to incorporate both human-driven and SAV shuttle dispatching and routing. Human-driven shuttles were modeled with delayed driver response times to trip requests while idling and were subject to safety \revision{protocols} that prohibited re-routing while en route between stops. SAVs enabled dynamic re-routing to accommodate on-demand trip requests while in operation and exhibited zero-delay responses to trip requests. Additionally, a MPC approach was developed for shuttle rebalancing toward designated idle locations without relying on extensive historical data, where two optimization models were proposed to determine optimal shuttle rebalancing strategies.

Scenario-based experiments were designed on varying SFLs, fleet sizes, and demand levels. A case study was implemented in the metropolitan Atlanta area using operational and ridership data from the MARTA Reach \revision{pilot}, the first pilot of ODMTS in Atlanta. \revision{The results demonstrated that incorporating SAVs in microtransit system operations can reduce user waiting times without compromising travel times while improving overall system efficiency through substantial reductions in total travel distance and empty kilometers driven.} The analysis of ridership and fleet size variations further illustrated how these benefits could scale between supply and demand. The cost analysis revealed that microtransit systems can yield cost savings with increasing demand.

Through this design framework, this study provides further justification for the viability of the ODMTS prototype and potential success of microtransit systems as a next-generation transit solution. While the MARTA Reach \revision{pilot} effectively connected riders to employment, healthcare, food, and other essential destinations, the integration of the technologies and operational strategies presented here could further enhance system performance.

\revision{One extension of this study is to account for behavioral responses under SAV operations. The MARTA Reach pilot data capture user behavior only in the context of human-driven microtransit systems and therefore do not provide observed behavioral responses to SAV services. User preferences and travel behavior may differ when interacting with SAV services. Future research could therefore extend the framework to capture these behavioral differences between human-driven and SAV operations. From an application perspective, future} research could examine the impact of electric vehicles and charging infrastructure within microtransit systems, which would introduce additional operational constraints related to shuttle range and charging availability. Additional work could also address the real-world challenges in integration of AVs as extensions to public transit systems. From a methodological perspective, further research may focus on incorporating multimodal transit incentive programs in microtransit systems to enhance user adoption \citep{lu2025incentive}. Further works on microtransit systems could connect other mobility services to integrate with the system as first- and last-mile connectors such as bikesharing and e-scooters \citep{lim2022bicycle, guan2025heuristic}.

\section*{Acknowledgments}
This research was partly supported by NSF CIVIC grant No. 2133342, NSF Leap-HI grant No. 1854684, and NSF grant No. 2046372. The authors would like to thank Dr. Jorge Huertas for providing benchmark visuals for this paper. The authors would like to thank MARTA for a successful collaboration.

\begin{appendices}
    \section{Vehicle Rebalancing Optimization with Tie-Breaking}
\label{appendix:rr-opt-w-tie-breaking}

Model~\eqref{formulation:vr-ratio} may produce ties in the ratio of mapped requests to shuttles at idle stops.
However, resolving ties prematurely in this model could lead to poor relocation decisions.

Consider a set of idle stops $\revision{X^\tau} \subseteq S_{idle}$ that exhibit ratio ties \revision{at epoch $\tau$}. In an example, suppose idle stops $s_a \in \revision{X^\tau}$ and $s_b \in \revision{X^\tau}$ are tied, and that each currently has two vehicles allocated. Suppose also that there are three vehicles located near stop $s_a$ and no vehicles near stop $s_b$. In this example, to minimize total vehicle relocation time, and thereby maximize vehicle availability, the tie-breaking procedure should allocate the majority of vehicles to stop $s_a$. This example illustrates two key principles: 1) tie-breaking only matters when $|\revision{X^\tau}|$ is greater than the remaining idle vehicles left to assign; and 2) that effective tie-breaking requires knowledge of individual vehicle locations. Thus, tie-breaking should occur during the assignment of vehicles for relocation.

Model~\eqref{formulation:tbrvr-mip} modifies  Model~\eqref{formulation:rvr-mip} to account for the relocation at tied idle locations. Constraint~\eqref{eq:tbrvr-mip_constr1} is analogous to constraint~\eqref{eq:rvr-mip_constr1} for idle locations that do not require tie-breaking. For all stops \revision{$s \in X^\tau$, either $\zeta_s^\tau$ or $\zeta_s^\tau+1$} vehicles could be assigned, as shown in constraints~\eqref{eq:tbrvr-mip_constr2} and~\eqref{eq:tbrvr-mip_constr3}. This flexibility allows more informed tie-breaking while minimizing total travel time.

\begin{mini!}
    %
    {}
    %
    {\eqref{eq:rvr-mip-obj},\label{eq:tbrvr-mip-obj}}
    %
    {\label{formulation:tbrvr-mip}}
    %
    {}
    %
    %
    \addConstraint
    {\sum_{v \in \revision{\hat{V}_{idle}^\tau}}     {\revision{w_s^{v,\tau}}}}
    {= \revision{\zeta_s^\tau} \quad \label{eq:tbrvr-mip_constr1}}
    {\forall s \in S_{idle} \setminus \revision{X^\tau}}
    \addConstraint
    {\sum_{v \in \revision{\hat{V}_{idle}^\tau}}     {\revision{w_s^{v,\tau}}}}
    {\geq \revision{\zeta_s^\tau} \quad \label{eq:tbrvr-mip_constr2}}
    {\forall s \in \revision{X^\tau}}
    \addConstraint
    {\sum_{v \in \revision{\hat{V}_{idle}^\tau}}     {\revision{w_s^{v,\tau}}}}
    {\leq \revision{\zeta_s^\tau} + 1 \quad \label{eq:tbrvr-mip_constr3}}
    {\forall s \in \revision{X^\tau}}
    \addConstraint
    {\eqref{eq:rvr-mip_constr2}, \eqref{eq:rvr-mip_binaryconstr}}
    {} 
    {\notag}
\end{mini!}%

\section{Additional Scenarios of Operational Costs}
\label{sec:additional-operational-costs}

Tables~\ref{tab:cost-per-rider-ridership-belvedere-additional} and~\ref{tab:cost-per-rider-ridership-westatlanta-additional} present additional sensitivity analysis of cost per \revision{user} at higher operational costs per shuttle per hour for Belvedere and West Atlanta, respectively.

\begin{table}[!ht]
\centering
\begin{tabular}{lrrr lrrr lrrr}

\toprule
\multicolumn{1}{c}{\multirow{2}{*}{\begin{tabular}[c]{@{}c@{}} Shuttle \\ Cost \end{tabular}}} & \multicolumn{3}{c}{$|N| = 39$: 1x (Base)}       &  & \multicolumn{3}{c}{$|N| = 78$: 2x}         &  & \multicolumn{3}{c}{$|N| = 117$: 3x}         \\ \cmidrule{2-4} \cmidrule{6-8} \cmidrule{10-12}
\multicolumn{1}{c}{} & $|V| = 3$      & $|V| = 4$     & $|V| = 5$      &  & $|V| = 3$      & $|V| = 4$     & $|V| = 5$    &  & $|V| = 3$      & $|V| = 4$     & $|V| = 5$     \\ \midrule
55                                                                                                             & 49.88     & 66.51     & 83.14     &  & 26.16       & 34.88     & 43.60   &  & 17.58     & 23.44    & 29.30   \\
60                                                                                                             & 54.42     & 72.56     & 90.70     &  & 28.54       & 38.05     & 47.56   &  & 19.18     & 25.57    & 31.97   \\
65                                                                                                             & 58.95     & 78.60     & 98.26     &  & 30.91       & 41.22     & 51.52   &  & 20.78     & 27.70    & 34.63   \\
70                                                                                                             & 63.49     & 84.65     & 105.81     &  & 33.29       & 44.39     & 55.49   &  & 22.38     & 29.84    & 37.30   \\
75                                                                                                             & 68.02     & 90.70       & 113.37       &  & 35.67     & 47.56        & 59.45    &  & 23.98        & 31.97    & 39.96   \\
80                                                                                                             & 72.56     & 96.74     & 120.93     &  & 38.05       & 50.73     & 63.41   &  & 25.57     & 34.10    & 42.62   \\
85                                                                                                             & 77.09     & 102.79     & 128.49     &  & 40.43       & 53.90     & 67.38   &  & 27.17     & 36.23    & 45.29   \\
90                                                                                                             & 81.63     & 108.84     & 136.05     &  & 45.18       & 60.24     & 75.30   &  & 28.77     & 38.36    & 47.95   \\
95                                                                                                             & 86.16     & 114.88     & 143.60     &  & 47.56       & 63.41     & 79.27   &  & 30.37     & 40.49    & 50.61   \\
100                                                                                                            & 90.70    & 120.93    & 151.16    &  & 50.00       & 66.67     & 83.33   &  & 31.97     & 42.62    & 53.28   \\ \hline
\end{tabular}
\caption{Cost (\$) per \revision{user} in Belvedere Across Ridership Levels. Shuttle costs are in \$/hour.}
\label{tab:cost-per-rider-ridership-belvedere-additional}
\end{table}

\begin{table}[!ht]
\centering
\begin{tabular}{lrrr lrrr lrrr}

\toprule
\multicolumn{1}{c}{\multirow{2}{*}{\begin{tabular}[c]{@{}c@{}} Shuttle \\ Cost \end{tabular}}} & \multicolumn{3}{c}{$|N| = 82$: 1x (Base)}       &  & \multicolumn{3}{c}{$|N| = 164$: 2x}         &  & \multicolumn{3}{c}{$|N| = 246$: 3x}         \\ \cmidrule{2-4} \cmidrule{6-8} \cmidrule{10-12}
\multicolumn{1}{c}{} & $|V| = 5$      & $|V| = 6$     & $|V| = 7$      &  & $|V| = 5$      & $|V| = 6$     & $|V| = 7$    &  & $|V| = 5$      & $|V| = 6$     & $|V| = 7$     \\ \midrule
55                                                                                                             & 40.17    & 48.20    & 56.24    &  & 19.64     & 23.57    & 27.50   &  & 13.24     & 15.89    & 18.54   \\
60                                                                                                             & 43.82    & 52.58    & 61.35    &  & 21.43     & 25.71    & 30.00   &  & 14.44     & 17.33    & 20.22   \\
65                                                                                                             & 47.47    & 56.97    & 66.46    &  & 23.21     & 27.86    & 32.50   &  & 15.65     & 18.78    & 21.91   \\
70                                                                                                             & 51.12    & 61.35    & 71.57    &  & 25.00     & 30.00    & 35.00   &  & 16.85     & 20.22    & 23.59   \\
75                                                                                                             & 54.78    & 65.73    & 76.69    &  & 26.79     & 32.14    & 37.50   &  & 18.06     & 21.67    & 25.28   \\
80                                                                                                             & 58.43    & 70.11    & 81.80    &  & 28.57     & 34.29    & 40.00   &  & 19.26     & 23.11    & 26.96   \\
85                                                                                                             & 62.08    & 74.49    & 86.91    &  & 30.36     & 36.43    & 42.50   &  & 20.46     & 24.56    & 28.65   \\
90                                                                                                             & 65.73    & 78.88    & 92.02    &  & 32.14     & 38.57    & 45.00   &  & 21.67     & 26.00    & 30.33   \\
95                                                                                                             & 69.38    & 83.26    & 97.13    &  & 33.93     & 40.71    & 47.50   &  & 22.87     & 27.44    & 32.02   \\
100                                                                                                            & 73.03    & 87.64    & 102.25   &  & 35.71     & 42.86    & 50.00   &  & 24.07     & 28.89    & 33.70   \\ \hline
\end{tabular}
\caption{Cost (\$) per \revision{user} in West Atlanta Across Ridership Levels. Shuttle costs are in \$/hour.}
\label{tab:cost-per-rider-ridership-westatlanta-additional}
\end{table}
\end{appendices}

\bibliographystyle{abbrvnat}
\bibliography{reference}
\end{document}